\documentclass[apj]{emulateapj}
\slugcomment{{\sc Accepted to ApJ:} June 13, 2016} 

\pdfoutput=1 

\usepackage{graphicx}
\usepackage{epstopdf}
\usepackage{multirow}
\usepackage{lscape}

\shorttitle{The Nature of Offset AGN}
\shortauthors{M\"uller S\'anchez et al.}

\begin{document}


\title{The Nature of Active Galactic Nuclei with Velocity Offset Emission Lines\footnotemark[1]
}


\author{F. M\"uller-S\'anchez$^{1}$, J. Comerford$^{1}$, D. Stern$^{2}$, F. A. Harrison$^{3}$}

\affil{$^1$ Department of Astrophysical and Planetary Sciences, University of Colorado, Boulder, CO 80309, USA}

%\affil{$^2$ Department of Physics and Astronomy, University of California, Los Angeles, CA 90095-1562, USA}
%\author{C. D. Biemesderfer\altaffilmark{4,5}}
%\affil{National Optical Astronomy Observatories, Tucson, AZ 85719}
%\email{aastex-help@aas.org}

\affil{$^2$ Jet Propulsion Laboratory, California Institute of Technology, 4800 Oak Grove Drive, Pasadena, CA 91109, USA}

%\affil{$^6$ Max Planck Institut f\"ur Radioastronomie, Auf dem Huegel 69, 53121, Bonn, Germany}

\affil{$^3$ California Institute of Technology, 1200 East California Boulevard, Pasadena, CA 91125, USA}

\footnotetext[1]{Based on observations at the W. M. Keck Observatory, which is operated as
a scientific partnership among the California Institute of Technology, the
University of California and the National Aeronautics and Space
Administration. The Observatory was made possible by the generous financial
support of the W. M. Keck Foundation.}

%\and

%\author{R. J. Hanisch\altaffilmark{5}}
%\affil{Space Telescope Science Institute, Baltimore, MD 21218}


%\altaffiltext{1}{Department of Physics, 366 Le~Conte Hall, University of 
%        California, Berkeley, CA, 94720-7300, United States}


\begin{abstract} 

We obtained Keck/OSIRIS near-IR adaptive optics (AO)-assisted integral-field spectroscopy to probe the morphology and kinematics of the ionized gas in four velocity-offset active galactic nuclei (AGNs) from the Sloan Digital Sky Survey. These objects possess optical emission lines that are offset in velocity from systemic as measured from stellar absorption features. 
%where the emission lines are offset in velocity from systemic
%, and one control AGN with emission lines that exhibit no velocity offset. 
At a resolution of $\sim0.18\arcsec$, OSIRIS allows us to distinguish which velocity offset emission lines are produced by the motion of an AGN in a dual supermassive black hole system, and which are produced by outflows or other kinematic structures. 
%We detected ionized gas emission in four velocity-offset AGNs. 
In three galaxies, J1018+2941, J1055+1520 and J1346+5228, the spectral offset of the emission lines is caused by AGN-driven outflows. In the remaining galaxy, J1117+6140, a counterrotating nuclear disk is observed that contains the peak of Pa$\alpha$ emission $0.2\arcsec$ from the center of the galaxy. The most plausible explanation for the origin of this spatially and kinematically offset peak is that it is a region of enhanced Pa$\alpha$ emission located at the intersection zone between the nuclear disk and the bar of the galaxy. In all four objects, the peak of ionized gas emission is not spatially coincident with the center of the galaxy as traced by the peak of the near-IR continuum emission. 
The peaks of ionized gas emission are spatially offset from the galaxy centers by $0.1\arcsec-0.4\arcsec$ ($0.1-0.7$ kpc). 
%the peak of continuum emission in $K-$band. 
We find that the velocity offset originates at the location of this peak of emission, and the value of the offset can be directly measured in the velocity maps. 
%In J$1354+1328$ (our control galaxy), the peak of ionized gas emission is spatially coincident with the center of the galaxy. While the emission line ratios of J$1354+1328$ can be well reproduced by AGN photoionization, 
The emission-line ratios of these four velocity-offset AGNs can be reproduced only with a mixture of shocks and AGN photoionization. Shocks provide a natural explanation for the origin of the spatially and spectrally offset peaks of ionized gas emission in these galaxies.

\end{abstract}

%\keywords{galaxies: active ---
%galaxies: nuclei --- 
%galaxies: kinematics and dynamics --- 
%infrared: galaxies}



\section{Introduction}\label{introduction}

%Galaxy mergers, which are a relatively common occurrence in hierarchical structure formation, can efficiently drive gas to the central galactic regions and fuel luminous quasars (e.g., Sanders & Mirabel 1996; Mihos & Hernquist

Galaxy mergers are a key component of galaxy evolution. When two galaxies collide, great amounts of gas can lose angular momentum and flow towards the center of the system. Mergers are believed to be responsible for triggering galaxy-wide starbursts, active galactic nuclei (AGN), and the creation of elliptical galaxies (e.g., Toomre 1977, Sanders et al. 1988, Genzel et al. 1998, Jogee 2004, Springel et al. 2005, Hopkins et al. 2006). As such, galaxy mergers have been the focus of much effort to understand the coordinated growth of supermassive black holes (SMBHs) and galaxies.
 
Because the majority of massive galaxies harbor central SMBHs (e.g., Kormendy \& Richstone 1995, Richstone et al 1998), galaxy mergers lead to the formation of SMBH pairs, and eventually, SMBH mergers and gravitational wave emission. This process can be divided into three distinct phases distinguished by the separation between the two SMBHs. During the early stages of a merger (at separations between 10 to several hundreds of kpc) the SMBHs reside at the centers of the progenitor galaxies. At separations between a few tens of parsecs (defined by the SMBH radius of influence) and 10 kpc, the SMBHs begin to spiral in toward the center of the merger system via dynamical friction (Begelman et al. 1980). At this stage they are best described as dual SMBHs, since their motion is governed by the galactic potential rather than their mutual gravity. When the two SMBHs become gravitationally bound to each other in a binary orbit, they form a SMBH binary system that eventually leads to the SMBH coalescence (Quinlan 1996, Jaffe \& Backer 2003). 

%and follow the motion of the whole system
%In the event of a merger between two massive galaxies, the supermassive black hole (SMBH) at the centre of each progenitor is expected to migrate towards the centre of the remnant (Begelman et al. 1980). When the two SMBHs become gravitationally bound to each other in a binary orbit, they continue to spiral inwards through the effect of three-body scattering of nearby stars (Quinlan 1996; Khan et al. 2011, 2013) until at small radii (. 0:01 pc) gravitational wave emission causes further shrinking of the orbit, eventually leading to the SMBHs coalescing (Jaffe&Backer 2003).

Although hierarchical galaxy formation predicts that dual SMBHs should be common, only a few are known. Many SMBHs are not active and therefore it is difficult to observe them directly using electromagnetic radiation. 
However, since gas-rich galaxy mergers trigger accretion-powered nuclear activity, and the AGN fraction increases as the SMBH separation decreases from $\sim80$ kpc to $\sim1$ kpc (Ellison et al. 2011, Koss et al. 2012), some dual SMBHs should be active. 
%Dual SMBHs are observationally identifiable when sufficient gas accretes onto them to power one or both SMBHs as AGN, in systems known as Òoffset AGNÓ and Òdual AGNÓ, respectively. 
When one or both SMBHs power AGNs, the systems are known as offset AGNs and dual AGNs, respectively\footnote{The term offset AGN refers to a single AGN in an on-going merger with nuclear separations less than 10 kpc. The majority of these AGNs are spatially offset from the dynamical center of the system (the only exceptions might be mergers between galaxies with fairly unequal masses or minor mergers). Note that the discovery of an offset AGN does not necessarily imply the presence of dual SMBHs, particularly in minor mergers, where the secondary nucleus might not contain a SMBH.}. 
%In other words, dual SMBH systems where only one is active.}. 

During the past few years, dual AGN candidates have been identified spectroscopically as galaxies with double-peaked narrow AGN emission lines (Comerford et al. 2009, Wang et al. 2009, Liu et al. 2010, Smith et al. 2010, Ge et al. 2012, Barrows et al. 2013, Comerford et al. 2013, Shi et al. 2014), and confirmations have been produced through follow-up observations at either radio or X-ray wavelengths (Komossa et al. 2003, Fu et al. 2011a, Koss et al. 2011, 2012, Mazzarella et al. 2012, Liu et al. 2013, Comerford et al. 2015, M\"uller-S\'anchez et al. 2015). 
However, these double peaks can also be produced by AGN-driven outflows and/or a rotating narrow line region (NLR; e.g., Heckman et al. 1981, Crenshaw et al. 2010). In fact, recent spectroscopic studies have found that most double-peaked AGNs are indeed produced by NLR kinematics (Shen et al. 2011, Fu et al. 2011b, Comerford et al. 2012, M\"uller-S\'anchez et al. 2015).

%Recently, 
Following a similar line of reasoning, Comerford \& Greene (2014) proposed that offset AGN candidates are identified spectroscopically as galaxies with narrow AGN emission lines that have a velocity offset from systemic (measured from the stellar absorption features). These are known as velocity-offset AGNs\footnote{This is a spectroscopic classification, similar to the term double-peaked AGNs. As such, they might or might not be offset AGN.}. In this scenario, the velocity offset is caused by a pair of SMBHs, consisting of an inactive SMBH and an active SMBH, which are 
displaced from the dynamical center of the host galaxy and inspiralling (Milosavljevi\'c \& Merritt 2001). 
%which is either inspiralling (Milosavljevi\'c \& Merritt 2001) or recoiling (Campanelli et al. 2007), and displaced from the dynamical center of the host galaxy. 
Comerford \& Greene (2014) identified 351 offset AGN candidates in the Sloan Digital Sky Survey (SDSS). However, as is the case for double-peaked narrow emission lines, the velocity offset can also be caused by gas kinematics such as outflows, inflows and a combination of disk rotation plus dust extinction (Crenshaw et al. 2010, Bae et al. 2014).    

Offset AGNs have so far received little attention despite their potential for yielding many more dual SMBH discoveries. 
This is largely due to the observational difficulty of detecting and resolving two galactic nuclei with separations $<10$ kpc. 
To date, there exists only one confirmed offset AGN in the galaxy NGC 3341 (Barth et al. 2008, Bianchi et al. 2013). Besides being spatially offset from the center of the galaxy by $\sim5.2$ kpc, this Seyfert 2 nucleus also exhibit a velocity offset of 200 km s$^{-1}$ from the systemic velocity. 
Two notable offset AGN candidates are Arp 220 (which probably hosts an AGN in the western nucleus; Engel et al. 2011, Teng et al. 2015), and Mrk 273 (which is either an offset AGN or a dual AGN; U et al. 2013). More recently, Allen et al. (2015) observed two velocity-offset AGNs identified by Comerford \& Greene (2014) as part of the Sydney-AAO Multi-object Integral field spectrograph (SAMI) Galaxy Survey. In both cases, their integral-field data did not provide evidence for the existence of an offset AGN. In the first galaxy, the authors concluded that the velocity offset is caused by a recent merger or gas accretion event (inflow), and in the second galaxy, an AGN-driven outflow produces the observed velocity offset. Although the SAMI data provide useful information regarding the nature of velocity-offset AGNs, the pixel size of the instrument ($0.5\arcsec$) is not ideally matched for the study of these galaxies, since the relevant scales are within an aperture of $3\arcsec$ diameter (the size of the SDSS fiber). 

In this work we use Keck Laser Guide Star Adaptive Optics (LGS-AO) with OH-Suppressing Infra-Red Imaging Spectrograph (OSIRIS) integral-field spectroscopy to further clarify 
%the ambiguities concerning 
the nature of AGNs with velocity offset emission lines. The $0.1\arcsec$ pixels of OSIRIS allow us to study in detail the spatial distribution and kinematics of the ionized gas, and distinguish which velocity offset emission lines are produced by the motion of an offset AGN, and which are produced by outflows or other kinematic structures.  
%For these 12 targets, OSIRIS observations will distinguish which offset AGN emission lines are produced by the motion of an AGN in a dual-SMBH system, and which are produced by outflows or
%gas kinematics. 
If the emission is produced by an offset AGN, the integral-field observations should show spatially offset compact 
emission in the flux maps of ionized gas with one consistent line-of-sight (LOS) velocity, where the velocity is equal to the velocity offset 
measured in the SDSS spectrum. This spatially and kinematically offset compact source of ionized gas should be surrounded by rotating gas that is settled in the gravitational potential of the galaxy. If the emission is produced by an AGN outflow, the integral-field observations should reveal a central AGN with spatially extended emission (and possibly secondary peaks of emission from the clouds in the NLR) with a wide range of blueshifted and redshifted velocities oriented along a direction opposite to the sense of rotation. Other typical signatures of outflows include broadened and/or asymmetric emission lines, very high LOS-velocities, and shock heating. The presence of some or all of these features would indicate an outflow rather than a dual SMBH origin for the observed velocity offset (see Section~\ref{outflows}). Other potential causes of the kinematic offset would result in different spatially resolved signatures; some specific cases are described in Section~\ref{identification}. 

This paper is organized as follows: In Section 2 we describe the sample characteristics, observations, and the methods used
for the data reduction and analysis. The general properties of the NLRs are described in Section 3. In this Section we also discuss the origin of the velocity offset AGN emission lines in each galaxy. Given that outflows are very common among velocity-offset AGN, in Section 4.1 we analyze in detail the galaxies containing AGN outflows and perform kinematic modeling. In addition, in order to explain the origin of the spatially offset peaks of emission in our flux maps, the possible contribution of shock-heating to photoionization is studied in Section 4.2. Finally, the overall conclusions of the study are outlined in Section 5. 

Throughout this paper, we use a flat $\Lambda$CDM cosmology with $\Omega_m=0.27$, $\Omega_\Lambda=0.73$, and $H_0= 70$ km s$^{-1}$ Mpc$^{-1}$.

%Another possibility is that the velocity offset is caused
%by a binary black hole system, consisting of an inactive
%black hole and an active black hole, which is either inspiralling
%(Milosavljevi«c & Merritt 2001) or recoiling (Campanelli
%et al. 2007), and displaced from the dynamical
%center of the host galaxy.

\begin{table*}
\caption[List of Velocity Offset AGN Observed with OSIRIS]{List of Velocity-Offset AGN Observed with OSIRIS.}
\begin{center}
{\small
\begin{tabular}{c c c c c c c c c}
\hline
\hline \noalign{\smallskip}
 SDSS Source & R.A. (J2000) & Dec. (J2000) & $z$ & kpc / $\arcsec$ & PA$_{\mathrm{obs}}$\tablenotemark{a}  & 
  t$_{\mathrm{int}}$\tablenotemark{b}  & Filter & Line Detected \\
 \hline
J$101847+294114$ & $10\,18\,47.57$ & $+29\,41\,14.10$ & $0.082$ & 1.8 & 100 & 30 & Kn1 & Pa$\alpha$ \\
J$105553+152027$ & $10\,55\,53.64$ & $+15\,20\,27.40$ & $0.092$ & 1.9 & 75 & 30 & Kn2 & Pa$\alpha$ \\
J$111519+542316$ & $11\,15\,19.98$ & $+54\,23\,16.70$ & $0.070$ & 1.5 & 180 & 30 & Kn1 & -- \\
J$111729+614015$ & $11\,17\,29.21$ & $+61\,40\,15.30$ & $0.112$ & 2.4 & 64 & 30 & Kn2 & Pa$\alpha$ \\
J$134640+522836$ & $13\,46\,40.79$ & $+52\,28\,36.60$ & $0.029$ & 0.5 & 140 & 30 & Jn3 & [Fe II] \\ 
%J$135429+132857$\tablenotemark{c} & $13\,54\,29.05$ & $+13\,27\,57.20$ & $0.063$ & 1.4 & 0 & 30 & Kn1 & Pa$\alpha$ \\
J$151858+204855$ & $15\,18\,58.17$ & $+20\,48\,55.30$ & $0.042$ & 0.8 & 144 & 30 & Jn4 & -- \\
\hline
\hline
\end{tabular}
}
\tablecomments{
$^\mathrm{a}$Position angle of OSIRIS observations, in degrees east of north. \newline
$^\mathrm{b}$Total integration time on source in minutes. 
}
%\tablenotetext{a}{Position angle of OSIRIS observations, in degrees east of north.}
%\tablenotetext{b}{Total integration time on source in minutes.}
%\tablenotetext{c}{This is the control galaxy in our sample (i.e., it does not show velocity offset emission lines).}
\end{center}
\label{table1}
\end{table*}

\section{Sample, Observations and Data Processing}\label{observations}

\subsection{Sample Selection}\label{sample}

The primary goal of this exploratory work is to identify bona fide offset AGNs among velocity-offset AGNs in the SDSS database (Comerford \& Greene 2014). 
%Our primary interest is to identify bona fide offset AGNs, which motivated our original search for offset AGN candidates among velocity-offset AGNs in the SDSS database (Comerford \& Greene 2014). 
Ideal offset AGN candidates have Seyfert-like optical spectra with narrow emission lines that exhibit a statistically significant ($>3\sigma$) LOS velocity offset relative to the host galaxy stellar absorption lines. They also have the velocity offsets of all their emission lines consistent to within $1\sigma$, which is expected for offset AGN but not
for the stratified velocity structure of outflowing AGN (e.g., Zamanov et al. 2002; Komossa et al. 2008), and show symmetric emission lines (multi-components or blueshifted/redshifted wings are characteristic of outflows, Crenshaw et al. 2010). Using these criteria, 
Comerford \& Greene (2014) identified 351 offset Type 2 AGN candidates. 
%with velocity offsets of 50 km s$^{-1} <  \left | \Delta v \right |< 410$ km s$^{-1}$.

We make additional cuts to select the targets most suitable for LGS-AO observations with OSIRIS. We select objects with large velocity offsets ($\left | \Delta v \right |> 50$ km s$^{-1}$) to select against miscentered fibers where rotating gas could
produce a slight redshifted or blueshifted offset in the emission lines. In addition, to optimize the use of
OSIRIS and obtain good S/N spectra, we select targets that are bright ($r < 16$), have luminous AGN emission (L$_{\mathrm{[OIII]}}>10^8$ L$_\odot$), and are at low redshift ($z < 0.12$). The low redshifts enable measurements of the smallest physical offsets of the AGN from the host galaxy
center (offsets $< 0.5$ kpc for our redshift range). 
%Finally, we select targets with sufficiently bright and nearby tip tilt stars to enable LGSAO observations.
These selection criteria yielded six offset AGN candidates in SDSS, which could be observed in one Keck night (Section~\ref{obs} and Table 1). 

%from which we could observe six in one Keck night (Section~\ref{obs} and Table 1). 
%In addition, we observed one active galaxy (J$1354+1328$) with narrow emission lines that do not exhibit a velocity offset relative to the host galaxy stellar absorption lines (Balmer lines have $\Delta v=35\pm17$ km s$^{-1}$, which is consistent with zero to within $3\sigma$). This control galaxy allows us to compare the characteristics of `normal' active galaxies in SDSS (spatial distribution, kinematics and line ratios) with those of velocity-offset AGNs. 

\begin{figure*}
\epsscale{.99}
\plotone{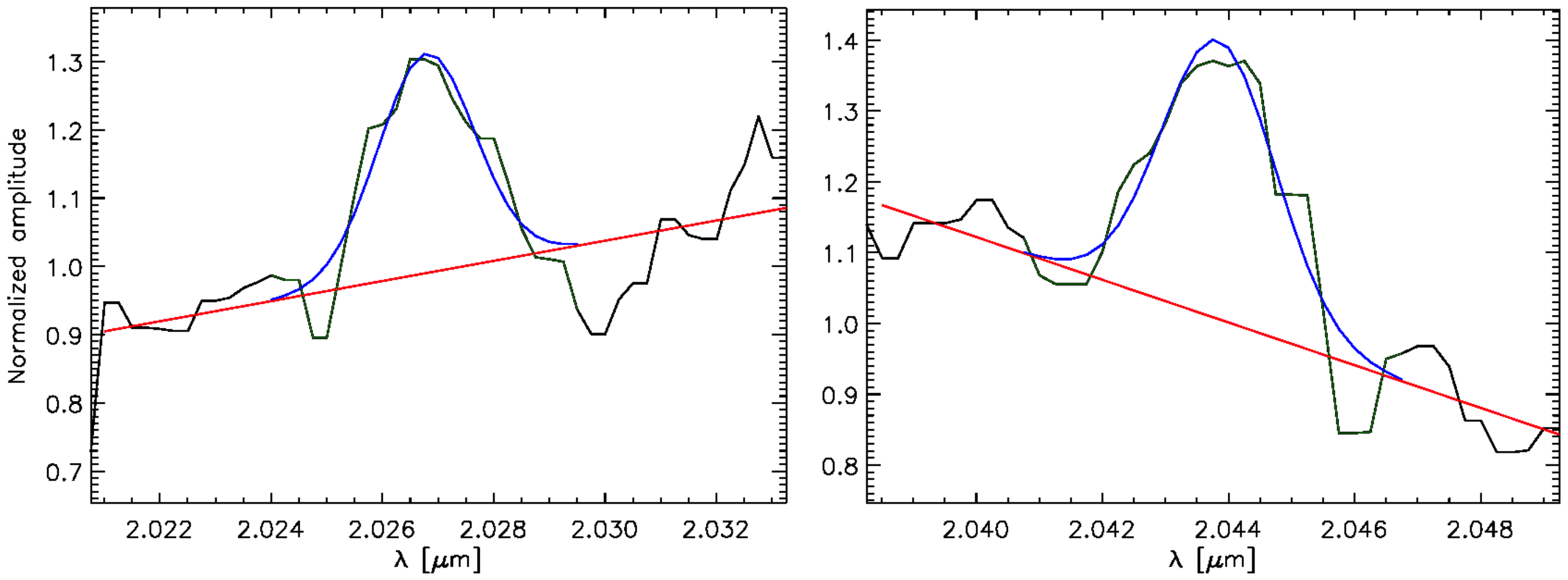}
\caption{Example Pa$\alpha$ line profiles extracted from the central regions of J$1018+2941$ (left) and J$1055+1520$ (right), chosen to represent the main types of profile shapes observed in the sample galaxies. The spectra have been extracted from apertures of $0.3\arcsec$ diameter. The data are shown by a solid black line, individual Gaussian fits by blue lines and the continuum level fit in red. 
%The green line indicates the spectral region where the emission line is observed. 
\label{fig1}}
\end{figure*}

\subsection{Observations and Data Reduction}\label{obs}

The six galaxies studied in this paper were observed on UT 2013 April 3 with LGS-AO and the integral-field spectrograph OSIRIS, commissioned at the W. M. Keck Observatory. OSIRIS  delivers spectra simultaneously over a contiguous two-dimensional (2D) field-of-view (FOV) in the wavelength range from 1.1 to 2.45 $\mu$m at a resolving power of $1500-4000$ (Larkin et al. 2006, van Dam et al. 2006, Wizinowich et al. 2006). In all cases we used the galaxy nucleus as the tip-tilt star, and the $0.1\arcsec$ pixel scale. This lenslet scale guarantees that our observations cover the entire area of an SDSS fiber (3$\arcsec$) for all the narrow filters of the instrument. We used Pa-$\alpha$ 1.87$\mu$m or Pa$\beta$ 1.26$\mu$m to trace the spatial distribution and kinematics of the ionized gas in the sample galaxies (preferably Pa$\alpha$, since it is the strongest recombination line of hydrogen in the near-IR). Given the redshift range of our targets, we used four different filters during that night. The average spectral resolution of our data is $R\sim3000$. 
%Since Pa-$\alpha$ is the strongest recombination line of hydrogen in the near-IR, we used the 
%We will observe the Pa? Given the redshift a field-of-view of $6.4\times4.2$, and a spectral resolution of R ? 3000 in the wavelength range 1.965Ð2.381 ?m. 
A summary of the data, including the specific exposure times and filters used for each galaxy, is given in Table 1.

The OSIRIS data were reduced using the OSIRIS data reduction pipeline (DRP), which performs all the usual steps needed to reduce near-IR spectra, but with additional routines for constructing the data cube. The reduction steps include bias and background subtraction, channel levels adjustments, crosstalk removing, glitch identification, cosmic ray cleaning, correction for nonlinearity, alignment and interpolation of the data, extraction of the spectra, wavelength calibration, sky subtraction, and co-addition of different exposures. Telluric correction and flux calibration were performed using a standard star (A- or B-type) observed near in time and close in airmass to the observations of the SDSS galaxies. We reduced the standard stars in the same way as the science frames. For each telluric star we extracted a one-dimensional spectrum, and divided this spectrum by a black body function with the same temperature as the star. Since the magnitude of the star is known, this yields a correction curve that can be applied to all further observed spectra for calibrating their fluxes. 
The uncertainties in the flux calibration are usually dominated by systematic errors in the observations, such as imperfect centering of the objects in the FOV and seeing variations (Erb et al. 2005). The errors caused by these effects can be estimated by comparing the fluxes received in each of the individual exposures that were co-added to produce the final data cube. We find that flux levels between exposures vary by about $\sim20\%$. 
%process are both substantial and difficult to quantify; however, we have attempted to estimate them in several ways. 
%First, we extracted $1\sigma$ error spectra for each of the galaxies; these primarily reflect the noise of the sky background. By integrating the flux in the variance spectrum (sigma^2) at the position of Pa_alpha and taking the square root of the result, we can measure the random error associated with the observation: this is $\sim10\%$. More difficult to measure are systematic errors: The largest sources of uncertainty are the flux lost due to imperfect centering of the objects on the slit, seeing and seeing variations. 
%We can get a sense of the importance of these effects by comparing the fluxes received in each of the individual exposures that were co-added to produce our final data cube. We find that flux levels between exposures vary by about $\sim30\%$; this includes random as well as systematic error. The uncertainty in the mean flux of our three or four exposures is then $15-20\%$. This accounts for variations in object centering and seeing. 
%We can perform a further check by calibrating the same object with several different standard stars; in doing sows find variations in flux of $15\%$ at maximum. 
We can perform a further check by comparing the photometric flux from 2MASS images with the continuum flux in $3\arcsec-5\arcsec$ apertures. Agreement between the different data sources was consistent to $15-20\%$. 
%because we have 2MASS photometry for all galaxies, we can compare the photometric flux with the continuum flux in 3-5" apertures. Based on all these tests, we take our measured fluxes as uncertain by about $15-25\%$. 

%In aXe flux calibration is done by dividing the uncalibrated spectrum that is in units of e-/s/pixel by a response function which has units of erg/s/cm2 /A ÐperÐ e-/s/pix, accounting for the width of each bin delta_lambda.

%We reduced the telluric stars in the same way as the science frames. For each telluric star we extracted a one-dimensional spectrum, removed the hydrogen Brackett ? absorption line at 2.166?m after fitting it with a Lorentzian profile, and divided the star spectrum by a black body spectrum with the same temperature as the star. Thus, when we divided each individual spaxel in the six galaxy data cubes by the corresponding best-fitting telluric spectrum, we also obtained a relative flux calibration.

As can be seen in the final column of Table 1, in four galaxies the ionized gas emission is detected at or above the desired S/N ratio (20). In the other two galaxies, the detection of the emission lines has been challenging due to the following reasons: in one case (J$1115+5423$), the emission line lies exactly at the bottom of the strongest telluric absorption in $K$-band, and in the other case (J$1518+2048$), the data were taken under suboptimal observing conditions in the last hour of the night (bad seeing, clouds, technical problems with the LGS system) so one of the frames has no data, and the remaining 20 minutes on-source do not have the required S/N to detect emission lines (Table 1). It is important to point out that in J$1346+5228$ Pa$\beta$ emission was not detected, but fortunately we could study the properties of the ionized gas using the [Fe II] line at rest wavelength $1.21$ $\mu$m, which was redshifted to the wavelength range of the Jn3 filter of OSIRIS. As a result, our subsequent analyses focus on the four galaxies for which the desired S/N was achieved (Table 2).

\begin{table*}
\caption[Summary of results]{Summary of results.}
\begin{center}
{\scriptsize
\begin{tabular}{l c c c c c c c c c c}
\hline
\hline \noalign{\smallskip}
 SDSS name & PA$_{\mathrm{gas}}$\tablenotemark{a} & PA$_{\mathrm{cont}}$\tablenotemark{b} & PA$_{\mathrm{gal}}$\tablenotemark{c} & $\Delta v_{\mathrm{SDSS}}$\tablenotemark{d} & $\Delta v_{\mathrm{Keck}}$\tablenotemark{e} & v$_{\mathrm{max}}$\tablenotemark{f} & 
 $\sigma_{\mathrm{mean}}$ &  $\sigma_{\mathrm{max}}$ & Kinematic & Origin of \\
  & ($\degr$E of N) & ($\degr$E of N) & ($\degr$E of N) & km s$^{-1}$ & km s$^{-1}$ & km s$^{-1}$ & km s$^{-1}$ & km s$^{-1}$ & components & the offset \\
 \hline
J$1018+2941$ & $8\pm10$ & $90\pm9$ & $99\pm7$ & $-51\pm14.5$ & $-55\pm12$ & $220\pm8$ & $140\pm15$ & $280\pm10$ & Outflow (D)\tablenotemark{g}  & Outflow \\
 & &  & & & & & & & rotation & \\
J$1055+1520$ & $135\pm11$ & $75\pm7$ & $75\pm6$ & $-113\pm14.7$ & $-120\pm15$ & $310\pm10$ & $150\pm17$ & $340\pm13$ & Outflow (D) & Outflow \\
J$1117+6140$ & $105\pm10$ & $110\pm8$ & $54\pm8$ & $85\pm16.6$ & $77\pm13$ & $180\pm7$ & $130\pm10$ & $230\pm10$ & large-scale disk & Inflow \\
& &  & & & & & & & nuclear disk & \\
J$1346+5228$ & $175\pm9$ & $140\pm7$ & $140\pm6$ & $-52\pm14.6$ & $-60\pm12$ & $580\pm11$ & $160\pm14$ & $300\pm14$ & Outflow (D) & Outflow \\ 
%J$1354+1328$ & $50$ & $60$ & $60$ & $35\pm17$ & $26\pm11$ & $150\pm10$ & $100\pm15$ & $200\pm15$ & Rotation (D) & --\\
%& &  & & & & & & & blueshifted tail & \\
\hline
\hline
\end{tabular}
}
\tablecomments{
$^\mathrm{a}$Position angle of the extended ionized gas emission. \newline
$^\mathrm{b}$Position angle of the near-IR continuum. \newline
$^\mathrm{c}$Position angle of the photometric major axis of the galaxy, from SDSS. \newline
$^\mathrm{d}$Velocity offset of the optical emission lines measured in SDSS spectra (measured from the Balmer lines). \newline
$^\mathrm{e}$Velocity offset of the Pa$\alpha$ emission line ([Fe II] in J$1346+5228$) measured in our OSIRIS spectra. \newline
$^\mathrm{f}$Maximum radial velocity measured in the OSIRIS spectra. \newline
$^\mathrm{g}$The letter D indicates the dominant kinematic component, except in J$1117+6140$, where the two kinematic components are dominant at different scales. See Section 4 for details. 
}
%\tablenotetext{a}{Position angle of the extended ionized gas emission.} %The typical error is $\sim10\degr$.}
%\tablenotetext{b}{Position angle of the near-IR continuum.} %Same error as in PA$_{\mathrm{gas}}$.}
%\tablenotetext{c}{Position angle of the photometric major axis of the galaxy, from SDSS.} %The typical error is $\sim7\degr$.}
%\tablenotetext{d}{Velocity offset of the optical emission lines measured in SDSS spectra (measured from the Balmer lines).}
%We followed a similar approach as in Comerford et al. (2012) to classify whether the AGN emission components in the long-slit spectra are spatially compact or extended. If both components appear spatially distinct and their FWHM in the spatial direction is comparable to the slit width ($W_s$) we label the object compact. If one or both of the components is slightly larger than the slit width (FWHM $> W_S$) or present asymmetric diffuse components (wings) in the spatial directions, we label the object extended.
%\tablenotetext{e}{Velocity offset of the Pa$\alpha$ emission line ([Fe II] in J$1346+5228$) measured in our OSIRIS spectra.}
%\tablenotetext{f}{Maximum radial velocity measured in the OSIRIS spectra}
%\tablenotetext{g}{The letter D indicates the dominant kinematic component, except in J$1117+6140$, where the two kinematic components are dominant at different scales. See Section 4 for details.}
\end{center}
\label{table2}
\end{table*}

\begin{figure*}
\epsscale{.99}
\plotone{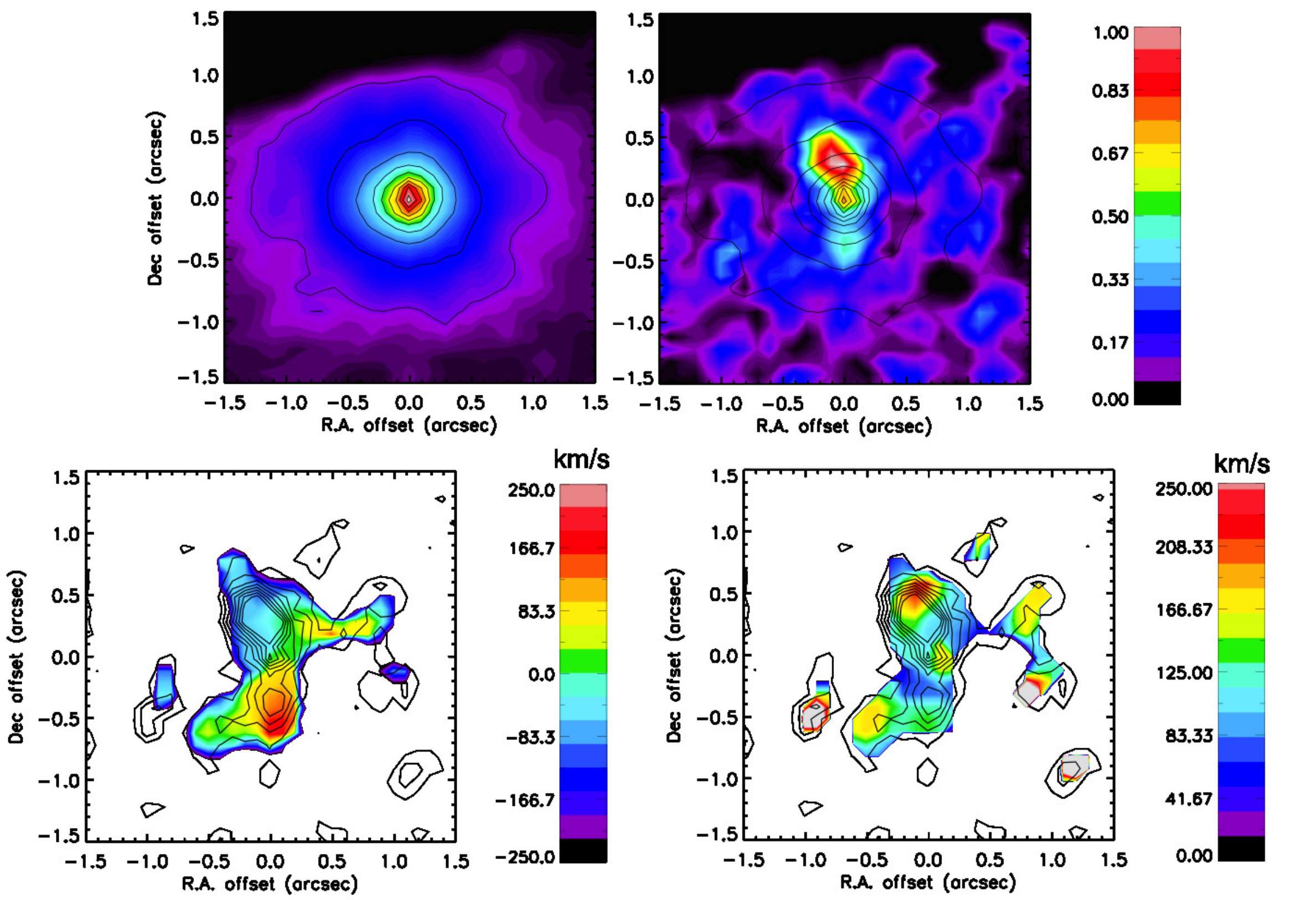}
\caption{Two-dimensional maps of the $K$-band continuum (top left) and Pa$\alpha$ flux distribution (top right), velocity (bottom left), and velocity dispersion (bottom right) for J$1018+2941$. The center of the galaxy (the peak of continuum emission at 1.9 $\mu$m) is located at position (0,0). The contours delineate the $K$-band continuum emission in the top panels, and the ionized gas morphology in the bottom panels. The continuum and flux maps are normalized to the peak of emission. 
The contours are set at 5, 10, 20, 30, 40, 50, 60, 70, 80 and $90\%$ of the peak of emission. Regions in white in the bottom panels correspond to pixels where the line properties are uncertain and thus were masked out. These rejected pixels in the velocity and dispersion maps are those with a flux density lower than 5\% of the peak of Pa$\alpha$ emission. In addition, very high velocity values (outside the chosen velocity scale) are not displayed in the image (white regions inside the contours). In all maps North is up and East is to the left.}
\label{fig2}
\end{figure*}

\subsection{Extraction of the Emission-line Gas Properties}\label{maps}

We derived 2D kinematic and flux distribution maps for each galaxy 
%emission-line flux distributions, radial velocity and velocity dispersion maps of the NLR 
using the IDL code LINEFIT. This
method, described in Davies et al. (2007a), fits the emission lines by convolving a Gaussian with a spectrally unresolved template profile (derived from telluric OH emission) to the continuum-subtracted emission-line profile at each spatial pixel in the data cube. The continuum is estimated using a polynomial function fitted to the line-free regions adjacent to the emission line. The continuum and line intervals must be set as input when executing the program. To allow for the best possible extraction of the line properties, the width of the fitting window is introduced manually after visual inspection of the line profile with the visualization tool QFitsView\footnote{See http://www.mpe.mpg.de/$\sim$ott/QFitsView/}. The fitting window should cover the complete spectral range of the line and enough range of continuum bluewards and redwards of the line. We estimate that the uncertainty in the continuum subtraction is on the order of $5\%$. This method allowed us to obtain uncertainties for the kinematic maps in the range of (5-20) km s$^{-1}$. The dispersion extracted by this fitting procedure is already corrected for instrumental broadening.

Despite some asymmetries observed in the emission-line profiles, we were successful in fitting a single Gaussian component
to the emission lines at each spaxel (Figure 1). The instrumental spectral profile is implicitly taken into account by convolving the assumed emission-line profile (the Gaussian) with a template line (tracing the effective instrumental resolution). As a result, small asymmetries in the emission-line profiles can be well matched. For more details on the key features of LINEFIT, see Davies et al. (2007a). %This sample represents the variety of different profile shapes we find across the integral field-unit fields and demonstrates the high quality of the data and the accuracy of the line fitting.

In addition, integrated intensity maps of Pa$\alpha$ ([Fe II]) were constructed by summing up all spectral channels containing line emission in each spatial pixel, and subtracting from each the continuum level. These measurements result in similar flux distributions as those obtained with LINEFIT, confirming that the single Gaussian fits result in good approximations for the fluxes. 
%the single Gaussian fits are good approximations.

Finally, we extracted an integrated spectrum in an aperture of $3\arcsec$ diameter for each galaxy. 
We find that the difference between the emission line velocities measured in our data and the stellar absorption line velocities measured in Comerford \& Greene (2014), $\Delta v_{\mathrm{Keck}}$, are consistent with the velocity offsets measured between the optical emission lines and stellar absorption lines ($\Delta v_{\mathrm{SDSS}}$, Table 2). 
%We find that the velocity offsets of the emission lines in our data, {\bf as measured from the stellar absorption features in the optical (Comerford \& Greene 2014), are consistent with the offsets measured in the SDSS spectra within the errors (see Table 2). 
However, some emission line asymmetries appear when adding the spectra in the $3\arcsec$ aperture, which are not present in the individual OSIRIS spectra and in the SDSS spectra of the sample galaxies. This will be further discussed in Section 4.3.

\section{Results}\label{results}

 \subsection{Morphology and Kinematics of the Ionized Gas}\label{general}

%In this Section, we discuss the emission-line properties of each object.

Figures $2- 5$ show the Pa$\alpha$ flux, velocity, and dispersion maps for the full galaxy sample, and an image of the $K-$band continuum (obtained directly from our Gaussian fits with LINEFIT). Note that in J$1346+5228$ the maps correspond to [Fe II] emission and $J-$band continuum. In the top panels, the contours delineate the near-IR continuum emission and the center of the galaxy, defined as the peak of continuum emission, is located at position (0,0). In the bottom panels, the contours delineate the morphology of the ionized gas and the regions in white correspond to pixels where the line properties are uncertain and thus were masked out. 
%These rejected pixels in the velocity and dispersion maps are those with a flux density lower than $5\%$ of the peak of Pa$\alpha$ ([Fe II]) emission.
These rejected pixels in the velocity and dispersion maps are those with a flux density lower than 5\% of the peak of Pa$\alpha$ ([Fe II]) emission. In addition, very high velocity values (outside the chosen velocity scale) are not displayed in the image (white regions inside the contours). 

Despite the diversity of the observed morphologies of the four galaxies, in each case most of the ionized gas emission consists of a bright peak of emission that is not located at the center of the galaxy %(except in J$1354+1328$, the control galaxy) 
and extended emission out to a few kpc. The emission in all cases is diffuse or filamentary, and in some cases it further breaks down into compact knots or blobs (J$1055+1520$ and J$1346+5228$) such as
those found in the NLRs of Seyfert galaxies (Schmitt et al. 2003, M\"uller-S\'anchez et al. 2011). 

We measured the extent of the photometric major axis of the ionized gas as well as its position angle (PA$_{\mathrm{gas}}$, Table 2). These measurements were done directly on the images, using as reference the contours corresponding to $5\%$ of the peak of Pa$\alpha$ ([Fe II]) emission. Using the same method, we also measured the photometric position angle of the near-IR continuum (PA$_{\mathrm{cont}}$, Table 2). A comparison of the measured PA$_{\mathrm{gas}}$ and PA$_{\mathrm{cont}}$ with measurements of the photometric position angle of the galaxy from SDSS images (PA$_{\mathrm{gal}}$) is presented in Table 2. In almost all cases, PA$_{\mathrm{cont}}$ is approximately equal to PA$_{\mathrm{gal}}$  (except in J$1117+6140$), and in three cases PA$_{\mathrm{gas}}$ is different from PA$_{\mathrm{cont}}$ (J$1018+2941$, J$1055+1520$, J$1346+5228$), indicating that in these galaxies the gas is probably not located in the galaxy disk.

As can be seen in the velocity maps of Figures $2-5$, there is a great deal of diversity in the kinematics of the ionized gas at these
scales in the sample galaxies. No clear rotational pattern can be identified.  
%(except in the control galaxy J$1354+1328$). 
The velocity maps of the four galaxies present redshifted and blueshifted velocities in the same direction as PA$_{\mathrm{gas}}$, and since in these galaxies PA$_{\mathrm{gas}}$ is different from PA$_{\mathrm{gal}}$ (Table 2), the direction of motion (radial velocity changing from blueshift to redshift) does not seem to follow ordered rotation in a disk. This is further discussed in Section~\ref{identification}.

The velocity dispersion maps, or $\sigma$-maps, show a wide range of values ranging from 30 km s$^{-1}$ to $\sim350$ km s$^{-1}$. Interestingly, in each galaxy, there exist regions where the velocity dispersion clearly increases. This is further discussed in Section~\ref{identification}. Table 2 lists the mean and the highest $\sigma$ measured for each of the galaxies.

\subsection{Identifying the Sources of Velocity Offset Emission Lines}\label{identification}

In this Section, we discuss the scenarios that may give rise to the velocity offset emission lines in our sample. 
%In order to draw general conclusions about the properties ofthe ionized gas in our sample galaxies, 
%all of the galaxies in the sample are evaluated using a consistent method to minimize differences due to spatial resolution and the shapes of the spectral profiles.
In the absence of kinematic data of the stars or the molecular gas for our galaxies, we assume in all cases that the orientation of the kinematic major axis of the galaxy disk is consistent with the position angle of the photometric major axis (PA$_{\mathrm{gal}}$). The results of the analysis are summarized in Table 2. 

%Four scenarios are considered: dual AGNs, radio jet-driven outflows, radiatively or mechanically driven AGN outflows, and disk rotation. %All these processes are related to AGN activity 
%In the four cases, the gas has been ionized by the AGN, as indicated by the measured optical line ratios, which are consistent with illumination by an AGN (Comerford et al. 2014). In other words, all these SDSS galaxies have AGN line ratios in both peaks. Therefore, a contribution from star formation is expected to be negligible. 

\begin{figure*}
\epsscale{.99}
\plotone{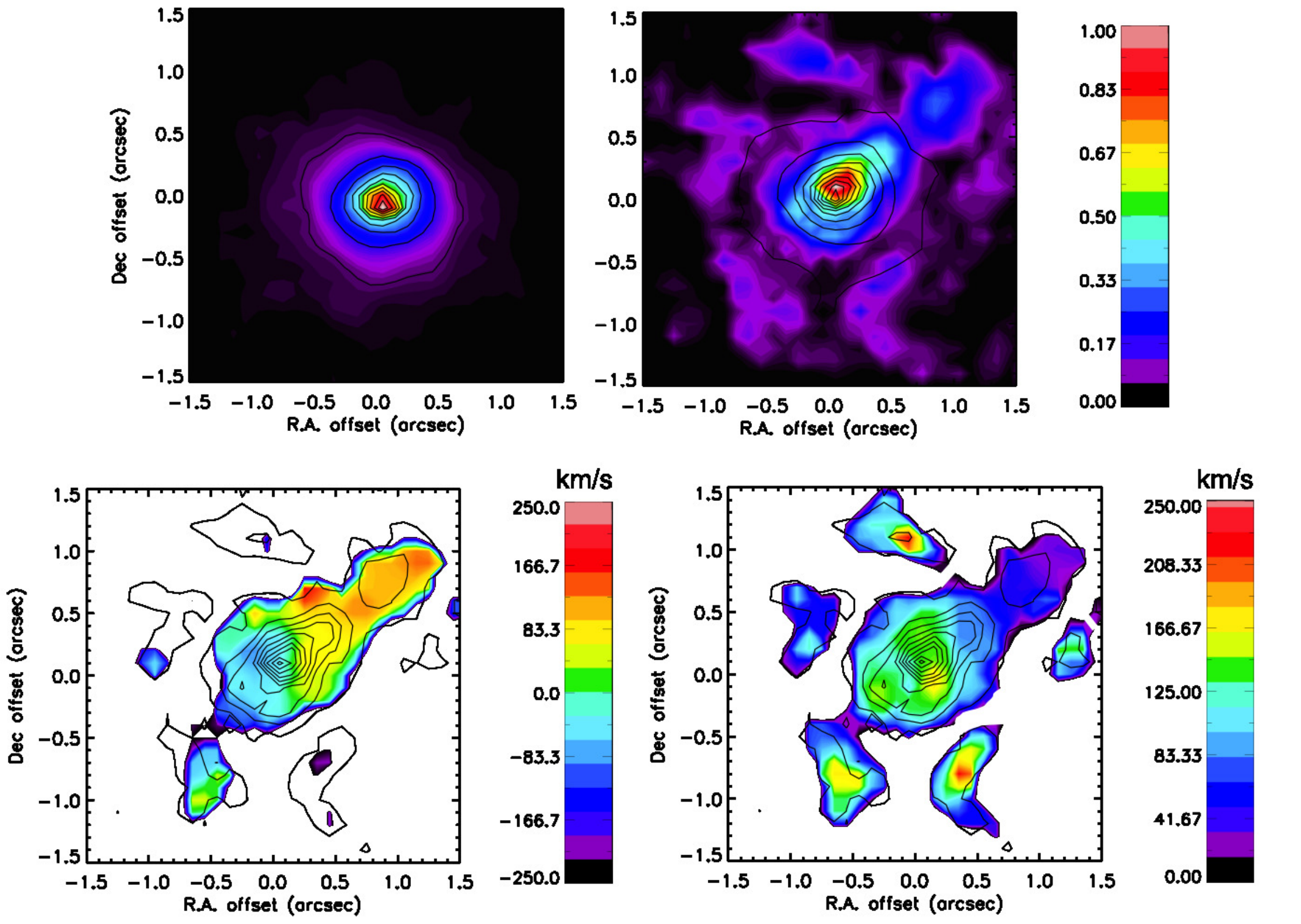}
\caption{Same as Figure 1 but for J$1055+1520$.
\label{fig3}}
\end{figure*}

\begin{figure*}
\epsscale{.99}
\plotone{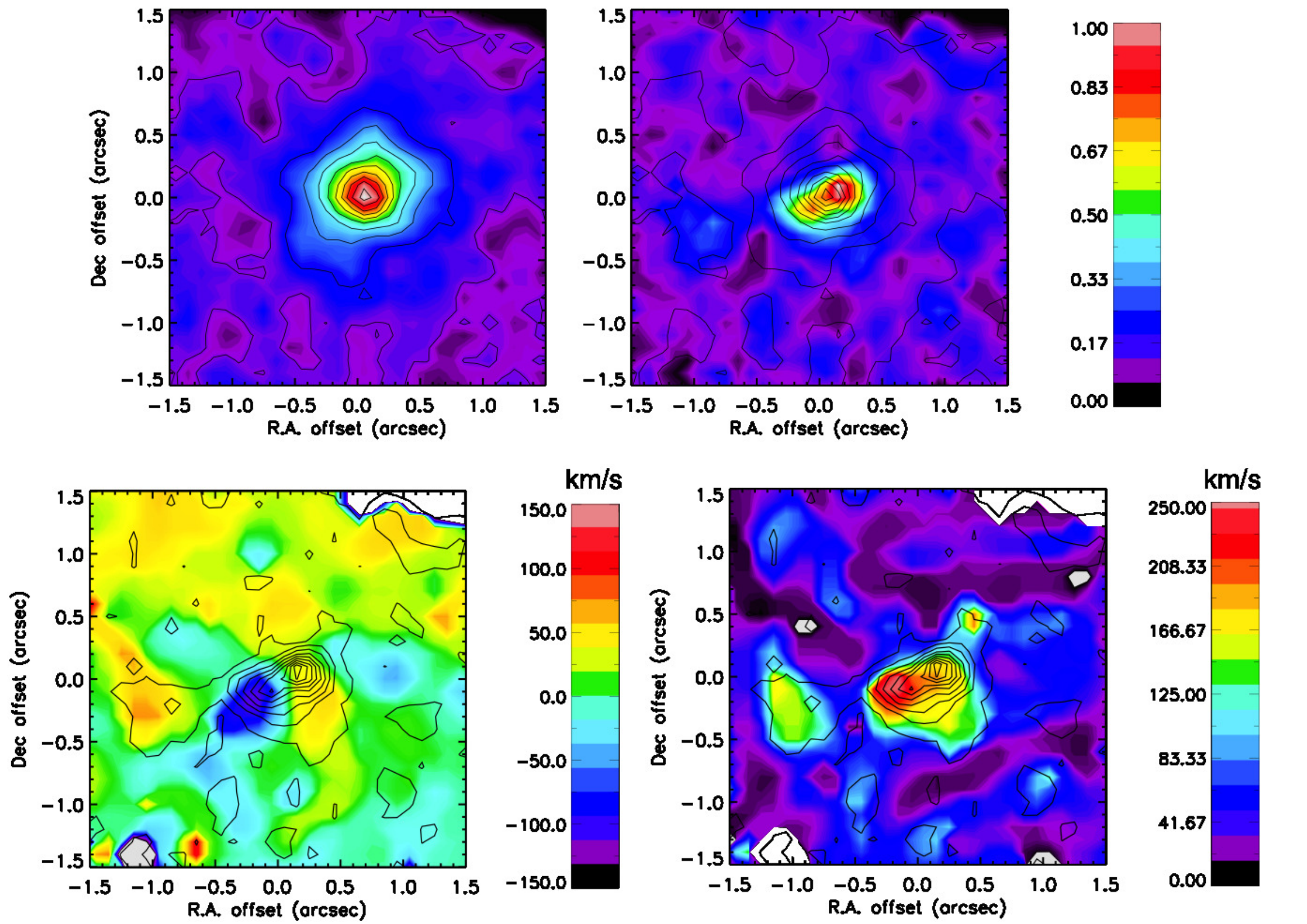}
\caption{Same as Figure 1 but for J$1117+6140$.
\label{fig4}}
\end{figure*}

\subsubsection{Velocity Offset Caused by AGN-driven Outflows}\label{outflows}

The spatial distribution and kinematics of the ionized gas in J$1018+2941$, J$1055+1520$ and J$1346+5228$ indicate the presence of outflows. Following M\"uller-S\'anchez et al. (2011), Fischer et al. (2013) and Harrison et al. (2014) among others, single-Gaussian components with LOS-velocities greater than 500 km s$^{-1}$ (detected in the continuum-subtracted maps with a significance of at least $3\sigma$) are direct evidence of outflows. Indeed, these velocities are too high to be explained by any sort of reasonable gravitational potential (including contributions from the SMBH, nuclear stellar cluster, and bulge; see also Rubin et al. 1985 and Das et al. 2007). In the absence of such strongly blueshifted or redshifted components, the adopted criterion for claiming the detection of an AGN-driven outflow is %that the following three conditions must be satisfied:
%the adopted criterion for claiming the detection of an AGN-driven outflow is 
that at least three of the following conditions must be satisfied: 

\begin{itemize}
	%\item Pa$\alpha$ emission with velocities greater than 500 km s$^{-1}$ is detected in the continuum-subtracted maps with a significance of at least $3\sigma$. 
	\item The spatial distribution of the ionized gas is extended in a different direction from the continuum emission (PA$_{\mathrm{gas}}\ne$PA$_{\mathrm{cont}}$).
	\item The velocity gradients of the gas (blueshifts and redshifts) are observed along a direction not coincident with the kinematic major axis of rotation as traced by PA$_{\mathrm{gal}}$. In other words, the ionized gas motions are extra-planar. A misalignment between the kinematical line of nodes and PA$_{\mathrm{gal}}$ should be observed (larger than the typical error in PA$_{\mathrm{gal}}$ of $\sim10\degr$, Table 2). 
	%\footnote{Outflows oriented along the plane of the galaxy do not satisfy criteria 1 and 2. In these cases}.
	\item The presence of broad emission line profiles (broader than forbidden lines from the NLR, typically
$\sim500$ km s$^{-1}$ FWHM; Osterbrock \& Pogge 1987, Taylor et al. 2003) in the individual spaxels of the data cube. 
	\item The velocity dispersion increases along the direction of the blueshifts and redshifts (PA$_{\mathrm{gas}}$).
	\item The velocity maps show evidence of radial acceleration to a projected distance, followed by deceleration to the systemic velocity. 
\end{itemize}

As can be seen in Table 2 and Figures 2, 3 and 5, J$1055+1520$ fulfills criteria $1-3$, J$1018+2941$ fulfills criteria $1-4$, and J$1346+5228$ fulfills all five criteria\footnote{Furthermore, this is the only galaxy in which spectral components with LOS-velocities  $>500$ km s$^{-1}$ are observed).}, indicating that in these three galaxies, the dominant kinematic component is radial outflow. A rotational component might be present in J$1018+2941$ and J$1346+5228$ as indicated by the kinematic minor axis (or zero velocity axis), which is nearly perpendicular to PA$_{\mathrm{gal}}$ in the central pixels, but the bulk of the gas is not rotating. A rotational component is not observed in J$1055+1520$. Additional evidence for outflows comes from the detection of asymmetric emission lines in the integrated spectra of these galaxies (Crenshaw et al. 2010, Section 4.3).

\subsubsection{J$1117+6140$: Velocity Offset Caused by Inflow into the Nuclear Region}\label{inflow}

Two kinematic components are observed in J$1117+6140$ at different scales (Figure 4): a disturbed rotational pattern is detected at larger scales, and in the central kpc, a counterrotating nuclear disk. An $HST$ image of this galaxy (GO 13513, PI: Comerford, Figure 6) reveals a bar extending along a position angle of $\sim110\degr$, which is consistent with PA$_{\mathrm{gas}}$ and PA$_{\mathrm{cont}}$, but is offset by $\sim40\degr$ from PA$_{\mathrm{gal}}$ (Table 2). This indicates that the bar is perturbing the kinematics of the large scale disk and is probably transporting gas to the nuclear region, creating a counter-rotating nuclear disk (PA$\sim85\degr$). 
%Inflow to the nuclear region nuclear bars has been observed in nearby Seyfert galaxies 
Steady-state circumnuclear inflow driven by a bar has been observed in local Seyfert galaxies (e.g., Fathi et al. 2005, Davies et al. 2014, Schnorr-M\"uller et al. 2016), and the formation of nuclear counter-rotating disks in spiral galaxies has been investigated in detail by Thakar \& Ryden (1996). The peak of Pa$\alpha$ emission is located in the redshifted part of the nuclear disk at $0.2\arcsec$ West from the center of the galaxy. The velocity offset of the emission lines in the SDSS spectrum originates at this position. The most plausible explanation for the nature of this peak is that it is a region of enhanced Pa$\alpha$ emission located at the intersection zone between the nuclear disk and the bar of the galaxy. Additional evidence for this interpretation comes from the presence of shocks in this region (Section 4.2).  

While the evidence for the bar in $HST$ images of J$1117+6140$ is strong, the velocity map of Pa$\alpha$ does not show clear kinematic evidence of the bar. This might appear in the velocity residuals after subtracting the two counter-rotating disks. To further investigate the inflow along the bar that settles in the nuclear disk, we model the Pa$\alpha$ kinematics as two rotating disks with kinemetry (Krajnovi\'c et al 2005). Briefly, kinemetry is an extension of surface photometry to the higher order moments of the velocity distribution. The procedure operates by first describing the data by a series of concentric ellipses  (or rings) of increasing major axis length, as defined by the system center, position angle, and inclination. At each radius, a small number of harmonic terms in a Fourier expansion is needed to determine the best fitting ellipse. If the motion is purely circular, the observed velocity along that ellipse is described by only one coefficient equivalent to the cosine law (Schoenmakers et al. 1997). 

For J$1117+6140$, the kinematic center is assumed to be coincident with the nucleus defined as the peak in the continuum emission. The velocity map is fit allowing the position angle (PA) and inclination angle ($q=$cos$i$) to vary. Subtraction of the rotational component to the data reveals the non-circular motions and these are investigated further for evidence of gas flow in a bar. The results can be seen in Figure 7. Kinemetry confirms the presence of two rotating disks at different scales. The PA of the concentric rings changes from $\sim90\degr$ in the central $0.7\arcsec$ to $\sim-145\degr$ at a radius $>1.2\arcsec$. These two PAs are consistent with the photometric PA of the nuclear disk and PA$_{\mathrm{gal}}$, respectively. Furthermore, there is a region between $0.8\arcsec-1.2\arcsec$ where the PA of the concentric rings changes from $90\degr$ to $\sim125\degr$. This is a transition region between the two disks and is probably influenced by the bar (PA$\sim110\degr$).  Interestingly, the two disks have the same inclination, as indicated by the approximately constant $q$ values which are in the range $0.4\arcsec-0.6\arcsec$ at all radii ($i=55\degr$ for a mean $q=0.57$, ignoring the first two pixels, i.e., $r<0.2\arcsec$). Therefore, the two disks are coplanar and is consistent with inflow driven by a large scale bar (which must be in the same plane as the two disks).  
The velocity residuals show two main components, blueshifted velocities along the bar and redshifted velocities almost perpendicular to the bar, both with velocities $\sim80$ km s$^{-1}$. Most of the two components might be interpreted in terms of $x_1$ and $x_2$ orbits. $x_1$ orbits are parallel to the large-scale bar (building up the straight gas lanes), and $x_2$ orbits are perpendicular to the bar (Contopoulos \& Grosbol 1989, Athanassoula 1992, van der Laan et al. 2011). Thus the observed velocity field indicates inflow in the bar that settles in the nuclear disk. 

%or an accreting SMBH that is not at the center of the galaxy, i.e. an offset AGN. 

%However, the nature of this peak is uncertain, with the two possibilities being a region of enhanced Pa$\alpha$ emission located at the intersection zone between the nuclear disk and the bar of the galaxy, or an accreting SMBH that is not at the center of the galaxy, i.e. an offset AGN. Follow-up observations with high spatial resolution at radio wavelengths would be useful to reveal the nature of this source.

%\subsubsection{J$1354+1328$: Control Galaxy with no Velocity Offset}\label{control}

%This galaxy does not show velocity offset emission lines in its SDSS spectrum, and therefore can be considered as the control object (Table 2). Two main differences are observed in this galaxy with respect to the other galaxies: the peak of Pa$\alpha$ emission is located at the center of the galaxy, and it is the only object in which ordered rotation in the plane of the galaxy is the dominant kinematic component (Figure 5). This spiral galaxy is an ongoing merger, with clearly visible tidal tails in optical images. It appears to be interacting with J$1354+1327$, which is located $10.2\arcsec$ (12.5 kpc) away from J$1354+1328$ and has a lower redshift (i.e., it is in front of J$1354+1328$ in our LOS). Therefore the blueshifted linear filamentary structure observed in the OSIRIS maps can be associated with a streamer of gas that is emanating from J$1354+1328$ and is traveling towards J$1354+1327$ with an average constant velocity of $69\pm 19$ km s$^{-1}$ (Figure 5). 

\begin{figure*}
\epsscale{.99}
\plotone{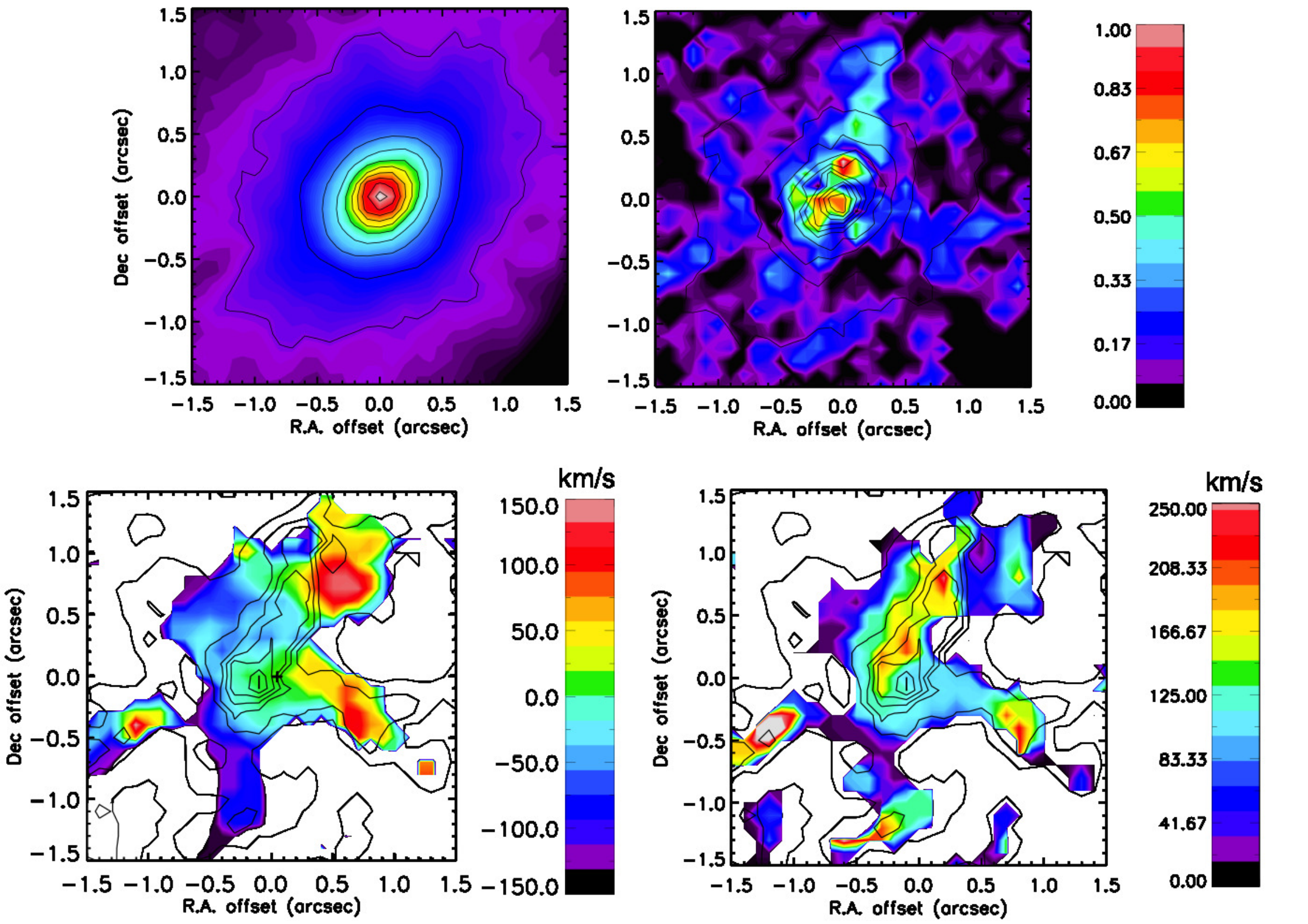}
\caption{Two-dimensional maps of the $J$-band continuum (top left) and [Fe II] flux distribution (top right), velocity (bottom left), and velocity dispersion (bottom right) for J$1346+5228$. The contours delineate the $J$-band continuum emission in the top panels, and the ionized gas morphology in the bottom panels. The center of the galaxy (the peak of continuum emission at 1.2 $\mu$m) is located at position (0,0). 
The continuum and flux maps are normalized to the peak of emission. 
The contours are set at 5, 10, 20, 30, 40, 50, 60, 70, 80 and $90\%$ of the peak of emission. Regions in white in the bottom panels correspond to pixels where the line properties are uncertain and thus were masked out. These rejected pixels in the velocity and dispersion maps are those with a flux density lower than 5\% of the peak of [Fe II] emission. In addition, very high velocity values (outside the chosen velocity scale) are not displayed in the image (white regions inside the contours). %Regions in white in the bottom panels correspond to pixels where the line properties are uncertain and thus were masked out. These rejected pixels in the velocity and dispersion maps are those with a flux density lower than 5\% of the peak of [Fe II] emission. The continuum and flux maps are normalized to the peak of emission. The units of the kinematic maps are km s$^{-1}$. 
In all maps North is up and East is to the left.
\label{fig5}}
\end{figure*}

\begin{figure}
\epsscale{.99}
\plotone{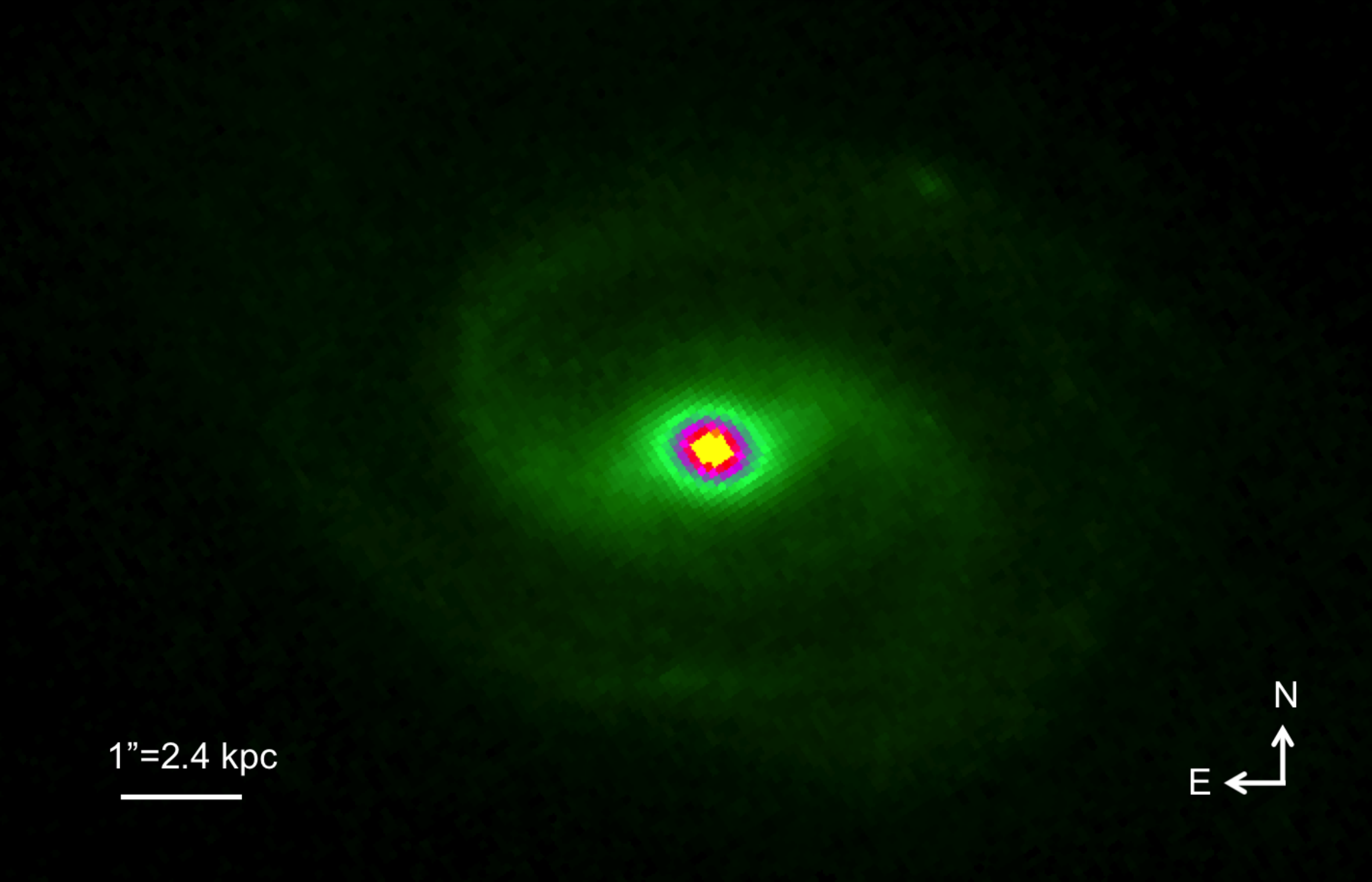}
\caption{$HST$ F160W image of J$1117+6140$. Note the large scale disk with a photometric PA of $\sim54\degr$, the bar extended along a PA of $\sim110\degr$ and the nuclear disk (PA$\sim85\degr$). 
\label{fig6}}
\end{figure}

\begin{figure*}
\epsscale{.99}
\plotone{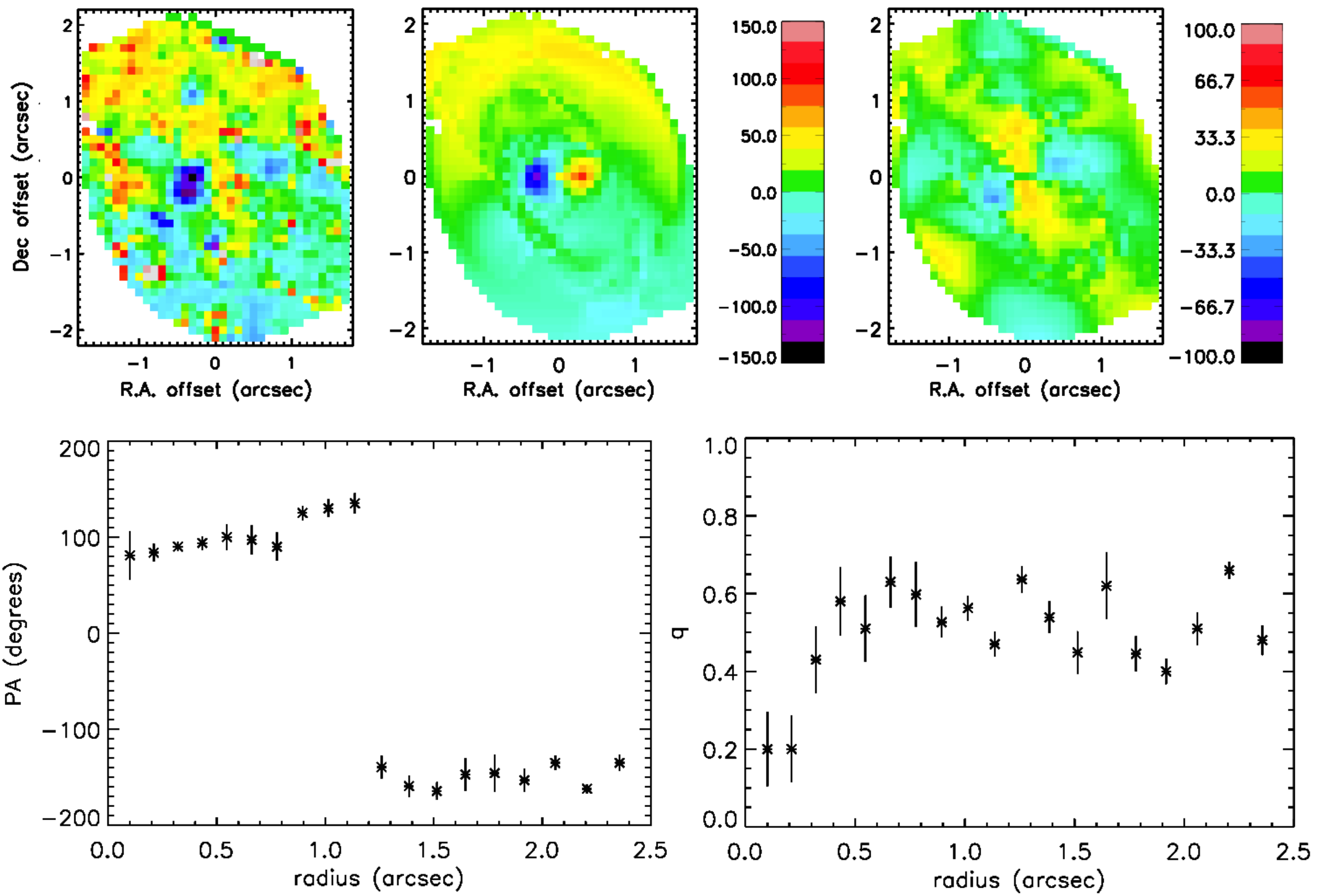}
\caption{Kinematic modeling of J$1117+6140$ with kinemetry. The maps are centered at the position of the $K-$band continuum. \textit{Top left:} Pa$\alpha$ velocity field used for the modeling; \textit{Top middle:} Best-fit kinematic model of two rotating disks; \textit{Top right:} Residuals (data - model) to the fit; \textit{Bottom left:} Kinematic PA; \textit{Bottom right:} Flattening $q=$cos$i$. Note the abrupt change in PA indicating the presence of two rotating disks.  
\label{fig7}}
\end{figure*}

\section{Discussion}

\subsection{Outflow Properties}\label{feedback}

Since three out of four velocity-offset AGN in our sample exhibit AGN-driven outflows of ionized gas, we investigate in this section the properties of these outflows and their impact on the evolution of their host galaxies.

Throughout our analysis we are interpreting our observations in the context of kinematic models of the NLR. Specifically, we are comparing our 2D kinematic data with 3D models of biconical outflows %superimposed on disk rotation 
(M\"uller-S\'anchez et al. 2011). A simple biconical geometry has been adopted to match the shape of the NLR as suggested by popular unification models of AGN
(e.g., Antonucci 1993, Urry \& Padovani 1995). The bicone model consists of two symmetrical hollow cones having interior and exterior walls with apices coincident with a central point source. Material flows inside the walls of the bicone, i.e. each point of the bicone corresponds to a velocity vector.
The simplest velocity law which reproduces the observations is radial linear acceleration followed by radial linear deceleration. To match the amplitudes of the measured blueshifted and redshifted velocities, the bicone can be tilted and rotated in 3D, and then projected in the plane of the sky. Thus, the velocity value of each element of the projected bicone corresponds to the average of the components of the velocity vectors in the direction of the LOS, as would be measured by the observer. The bicone model has seven free parameters: inner and outer half-opening angles ($\theta_{\mathrm{in}}$ and $\theta_{\mathrm{out}}$), position angle and inclination of the bicone's major axis (PA$_{\mathrm{cone}}$ and $i_{\mathrm{cone}}$), height of each cone ($z_{\mathrm{max}}$), turnover radius at which the deceleration phase starts ($r_t$) and maximum velocity ($v_{\mathrm{max}}$). Additional details of the modeling can be found in M\"uller-S\'anchez et al. (2011). 

%The disk model has three free parameters: magnitude of the circular velocity ($v_c$), position angle of the major axis of the disk (P.A.$_{\mathrm{disk}}$), and inclination of the disk ($i_{\mathrm{disk}}$). As can be seen in Figure 1 for the case of NGC 3081, our kinematic models of biconical outflow plus rotation provide a good match to the observed velocity fields of the NLR/CLR (see also \cite[M\"uller-S\'anchez et al. 2011]{mueller_etal11}). 

The results of the kinematic modeling are shown in Figures 8, 9 and 10 for J$1018+2941$, J$1055+1520$ and J$1346+5228$, respectively. 
%As can be seen in Figure~\ref{fig6} for the case of J$1018+2941$ , 
As can be seen in these Figures, our kinematic models of biconical outflows provide a good match to the observed velocity fields of the NLR. No obvious regular patterns are noticeable in the residuals maps, suggesting that the features observed in the residuals are dominated by noisy pixels in the data or other unresolved kinematic substructure. Furthermore, the residuals show mostly values between $-20$ to $20$ km s$^{-1}$, consistent with the measured errors of the velocity maps. The results of applying this method to the three galaxies with outflows are summarized in Table 3. The biconical structures have full opening angles ($2\theta_{\mathrm{out}}$) in the range $70-110^\circ$. The typical distance at which the outflows reach $v_{\mathrm{max}}$ is $r_t=1.4$ kpc (with values ranging from $0.8-2.1$ kpc). The inclination of the bicone axis in the three cases is $<30\degr$ (nearly perpendicular to the LOS), suggesting that the equatorial plane of the central engine and the torus are viewed nearly edge-on. This is consistent with unification models, which predict a large inclination of the torus in Type 2 AGN (see e.g., Antonucci 1993, Davies et al. 2007b). 

From the measured geometrical parameters and velocities we estimate mass outflow rates ($\dot{M}_{\mathrm{out}}$) and kinetic power of the outflows ($\dot{E}_{\mathrm{out}}$) using the method described in M\"uller-S\'anchez et al (2011) and assuming a gas density $N_e=100$ cm$^{-3}$ at a typical turnover radius of 1 kpc (Bennert et al. 2006, Riffel et al. 2009, Schnorr-M\"uller et al. 2016). The filling factor $f$ is inversely proportional to $N_e$ and can be directly obtained from the Pa$\alpha$ luminosity as described in Riffel \& Storchi-Bergmann (2011) (see also Osterbrock \& Ferland 2006). We found $f$ values of 0.02 and 0.04 for J$1018+2941$ and J$1055+1520$, respectively, and adopted a typical value of $f=0.03$ for J$1346+5228$ (note that this is the galaxy where we only detected [Fe II] emission). 
%the three galaxies in our sample with outflowing NLRs.} 
%and a filling factor $f=0.01$ for the NLR (Bennert et al. 2006). 
The resulting mass outflow rates are in the range $72-630$ M$_\odot$ yr$^{-1}$ and the estimated kinetic powers between $2-19\times10^{42}$ erg s$^{-1}$. These are $\sim10^2$ times smaller than the AGN's bolometric luminosities, but they are of the order of the power required by feedback models to explain fundamental galaxy properties ($\sim0.005-0.05\, L_{\mathrm{bol}}$; di Matteo et al. 2005, Hopkins \& Elvis 2010). This means that these three velocity-offset AGNs have outflows that likely have significant effects on the evolution of the host galaxy.

Interestingly, the mass outflow rates measured in these velocity-offset AGNs are approximately an order of magnitude larger than those measured in local Seyfert galaxies (M\"uller-S\'anchez et al. 2011, Storchi-Bergmann et al. 2010). This is mainly due to the fact that $r_t$ (the distance from the nucleus at which the outflows reach maximum velocity) is larger in these objects than in Seyfert galaxies. We measure $r_t$ values in the range $1-3$ kpc. In contrast, for local Seyfert galaxies, several authors (e.g., Crenshaw et al. 2010, M\"uller-S\'anchez et al. 2011, Fischer at el. 2013) find turnover radii of a few hundred parsecs. The other geometrical parameters of the outflows and $v_{\mathrm{max}}$ are similar for these velocity-offset AGNs and Seyfert galaxies. 
There are two main differences in the velocity-offset AGNs from SDSS studied here when compared to local Seyfert galaxies: they are at slightly higher redshifts ($0.029<z<0.112$), and have slightly higher bolometric luminosities ($10^{43.5}<L_{\mathrm{bol}}<10^{45}$ erg s$^{-1}$)\footnote{In contrast, most local Seyfert galaxies are located at redshifts $z<0.025$ and have bolometric luminosities between $10^{42}$ and $10^{44}$ erg s$^{-1}$ (Schmitt et al. 2003, Hainline et al. 2013).}. This indicates that more luminous AGNs produce larger outflows and is consistent with the relationship found between AGN bolometric luminosity and the size of the NLR (see e.g., Bennert et al. 2006, Hainline et al. 2013, M\"uller-S\'anchez et al. 2015). However, although these velocity-offset AGNs possess larger, and thus more powerful outflows than Seyfert galaxies, the net effect  of the outflows in the host galaxies (in terms of feedback) appears to be the same in both type of objects. This is due to the fact that the ratio $\dot{E}_{\mathrm{out}}$/L$_{\mathrm{bol}}$ is the same for low luminosity (Seyfert galaxies) and moderate luminosity (the velocity-offset AGNs from SDSS studied here) AGN (between 0.005 and 0.05, M\"uller-S\'anchez et al. 2011). 

When comparing the outflows identified in this work with the outflows present in other SDSS galaxies at similar redshifts and luminosities (e.g., Harrison et al. 2014), we find similar outflow sizes and energetics. Harrison et al. (2014) measured outflows radii between 1.5 and 4.3 kpc and $\dot{E}_{\mathrm{out}}$/L$_{\mathrm{bol}}$ ratios of $0.005-0.1$. 
% (44 < logLbol < 46)
Note that the galaxies studied by Harrison et al. (2014) do not exhibit emission lines with significant velocity offsets as the ones selected for this work, but most of them contain powerful outflows. A larger sample of outflows measured in velocity-offset AGNs  would be needed to explore any differences between the outflows found in these objects and active galaxies having a near-to-zero-velocity offset. 

%the fact that these objects possess larger, and thus more powerful, AGN outflows does not imply that the effects of feedback are stronger in these galaxies.  

%The selected velocity-offset AGNs from SDSS have two main differences studied here are at slightly higher redshifts than the local sample of Seyferts and have slightly higher bolometric luminosities.  

\begin{table}
\caption{Kinematic model of the NLR %in J$1018+2941$, J$1055+1520$ and J$1346+5228$
}
\begin{center}
{\scriptsize
\begin{tabular}{l c c c}
\tableline\tableline
Parameter & J$1018+2941$%\tablenotemark{a} 
& J$1055+1520$ & J$1346+5228$ \\
\tableline
$z_{\mathrm{max}}$ (kpc) & $3.1^{+1}_{-0.9}$ & $5.3^{+1.8}_{-1.3}$ & $2.2^{+0.9}_{-0.6}$\\ 
$r_t$\tablenotemark{b} (kpc) & 1.2$^{+0.07}_{-0.06}$ & $2.1^{+0.12}_{-0.13}$ & $0.8^{+0.08}_{-0.07}$\\
$\theta_{\mathrm{in}}$ ($\degr$) & 18$^{+11}_{-7}$ & 27$^{+10}_{-7}$ & 30$^{+12}_{-10}$ \\
$\theta_{\mathrm{out}}$ ($\degr$) & 35$^{+5}_{-9}$ & 48$^{+9}_{-8}$ & 54$^{+10}_{-9}$  \\
$i_{\mathrm{cone}}$ ($\degr$) & 22$^{+8}_{-6}$ & 18$^{+9}_{-7}$ & 15$^{+10}_{-7}$ \\
PA$_{\mathrm{cone}}$ ($\degr$) & $10\pm7$ & $138\pm6$ & $120\pm7$ \\
$v_{\mathrm{max}}$\tablenotemark{a} (km s$^{-1}$) & 290$\pm15$ & 310$\pm20$ & $340\pm23$\\
$\dot{M}_{\mathrm{out}}$ (M$_\sun$ yr$^{-1}$) & $72\pm5$ & $630\pm45$ & $85\pm4$ \\
$\dot{E}_{\mathrm{out}}$ (10$^{42}$ erg s$^{-1}$) & $2.0\pm0.3$ & $19\pm4$ & $3.1\pm0.4$ \\
$\dot{E}_{\mathrm{out}}/L_{\mathrm{bol}}$ & $0.01\pm0.002$ & $0.06\pm0.015$ & $0.01\pm0.002$ \\
%$v_c$\tablenotemark{d} (km s$^{-1}$) & 60$\pm20$ \\
%$i_{\mathrm{disk}}$ ($\degr$) & 53$^{+10}_{-9}$ \\
%PA$_{\mathrm{disk}}$ ($\degr$) & 27$\pm9$ \\
%$\alpha$ \tablenotemark{e} ($\degr$) & 7 \\
%$\beta$ \tablenotemark{f} ($\degr$) & 55 \\
%$\big(\chi^2_{\mathrm{rot}}\big)_{\mathrm{min}}$\tablenotemark{g} & 423 \\
%$\big(\chi^2_{\mathrm{out}}\big)_{\mathrm{min}}$\tablenotemark{h} & 537 \\
%$\big(\chi^2_{\mathrm{rot+out}}\big)_{\mathrm{min}}$\tablenotemark{i} & 390 \\
%$N$\tablenotemark{j} & 330 \\
\tableline
\tableline
\end{tabular}
}
\tablecomments{The quoted parameters correspond to the model presenting the smallest $\big(\chi^2\big)_{\mathrm{min}}$ and the uncertainties refer to $3\sigma$ confidence level. The derived values are also the best-fit parameters for the NLR. \newline
$^\mathrm{a}$Full module of the velocity vector, i.e. relative to the nucleus not to the observer. \newline
$^\mathrm{b}$Radial distance from the AGN to the location where $v_{\mathrm{max}}$ is observed. 
}
%\tablenotetext{a}{No deceleration observed.}
%\tablenotetext{a}{Full module of the velocity vector, i.e. relative to the nucleus not to the observer.}
%\tablenotetext{b}{Radial distance from the AGN to the location where $v_{\mathrm{max}}$ is observed.}
%\tablenotetext{d}{Rotational velocity measured at $r=50$ pc.}
%\tablenotetext{e}{Angle between the outer edge of the bicone and the LOS.}
%\tablenotetext{f}{Relative tilt of the bicone axis to the disk.}
%\tablenotetext{g}{Minimum $\chi^2$ value for models incorporating only rotation.}
%\tablenotetext{h}{Minimum $\chi^2$ value for models incorporating only biconical outflow.}
%\tablenotetext{i}{Minimum $\chi^2$ value for models incorporating rotation and biconical outflow.}
%\tablenotetext{j}{Number of data points in the velocity field, equivalent to the number of degrees of freedom.}
\end{center}
\label{table3}
%\tablecomments{The quoted parameters correspond to the model presenting the smallest $\big(\chi^2\big)_{\mathrm{min}}$ and the uncertainties refer to $3\sigma$ confidence level. The derived values are also the best-fit parameters for the NLR. }
\end{table}

\begin{figure*}
\epsscale{.99}
\plotone{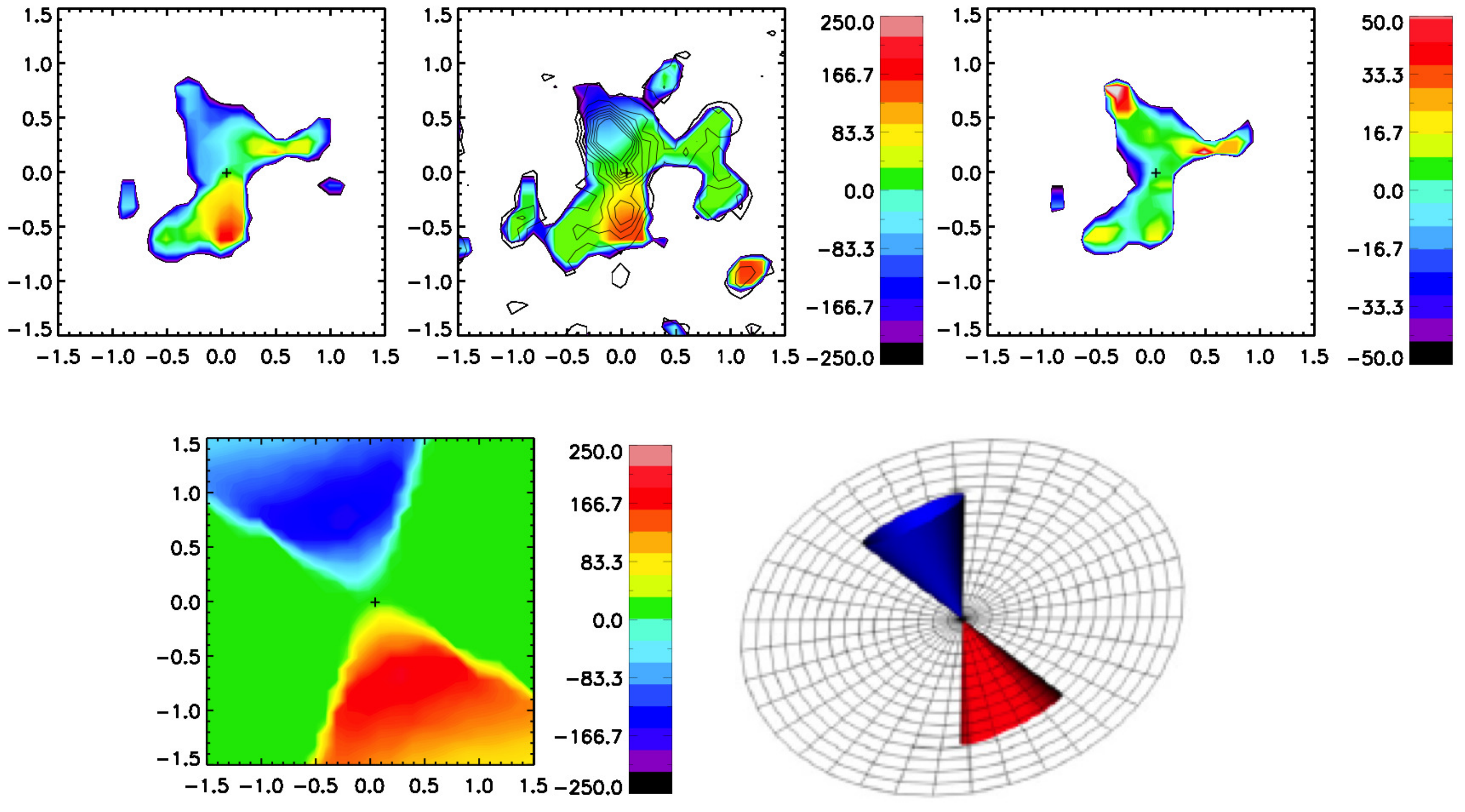}
\caption{Maps showing the data and the best-fit kinematic model used to reproduce the Pa$\alpha$ kinematics in J$1018+2941$. The maps are centered at the position of the peak of $K-$band continuum, and the scales are in km s$^{-1}$. \textit{Top Left:} Pa$\alpha$ velocity field used for the modeling (same as the one shown in Figure 2). \textit{Top Middle:} Best-fit kinematic model of a biconical outflow with flux contours. \textit{Top Right:} Residuals (data-model) to the fit. \textit{Bottom Left:} Bicone projection in the plane of the sky. \textit{Bottom Right:} Geometric model of the NLR and the inner galactic disk. 
\label{fig8}}
\end{figure*}

\subsection{Spatially Offset Peaks of Ionized Gas Emission: The Role of Shocks}\label{shocks}

The four velocity-offset AGN studied here all exhibit spatially offset peaks of ionized gas emission. In each case, the velocity of this peak corresponds to the velocity offset measured in the SDSS spectrum, and its value can be clearly seen in the velocity map (Figures 2 to 5). 
%In contrast, our control galaxy J$1354+1328$ does not exhibit secondary peaks of emission in the Pa$\alpha$ flux map and the velocity of the central peak is consistent with the systemic velocity of the galaxy (Figure 5). 

Spatially offset peaks of emission can be explained by the presence of shocks\footnote{We do not consider peaks caused by star-forming rings/nuclear star clusters since the optical line ratios of the galaxies in our sample are consistent with AGN-dominated emission rather than star formation (Comerford \& Greene 2014).}. Gas can be shocked at the interface between two kinematically distinct gas clouds, enhancing the emission at this location. In J$1018+2941$, J$1055+1520$ and J$1346+5228$, outflowing clouds drive shocks into the surrounding material, with an environmental impact that may be enough to produce peaks of ionized gas emission at different locations along the NLR (see e.g., Mazzalay et al. 2013, Meyer et al. 2015). The projected spatial offsets are approximately $0.38\arcsec$ ($\sim0.7$ kpc), $0.1\arcsec$ ($\sim0.2$ kpc), and $0.27\arcsec$ ($\sim0.15$ kpc) in J$1018+2941$, J$1055+1520$ and J$1346+5228$, respectively. In $1117+6140$, the peak of Pa$\alpha$ emission is offset by $\sim0.2\arcsec$ ($\sim0.5$ kpc) and may be the product of cloud-cloud collisions occurring at the interface between the bar and the nuclear disk. 

In order to explain the origin of the spatially offset peaks of emission in our flux maps, the possible contribution of shock-heating to photoionization must be examined. The effects of shocks that leave a signature on emission-line spectra can be summarized as follows (Dopita \& Sutherland 1996, Moy \& Rocca-Volmerange 2002): 
(i) shocks compress the gas increasing its density and temperature (up to millions of K), (ii) this high temperature gas emits both X rays and ultraviolet (UV) light, which contribute to the ionization in front of the shock and behind it, (iii) at the same time, compression locally reduces the ionization parameter, and thus enhances the low ionization emission lines such as [O II], [N II], [S II], and [O I], (iv) in the region between the X-ray emitting zone and the low excitation zone, the temperature decreases to $10^4-10^5$ K, boosting the emission of UV and optical collisional excited lines. 
%(i) shocks generate compression and thus high gas densities, (ii) this compression locally reduces the ionization parameter, and thus enhances the low ionization emission lines such as [O II], [N II], [S II], and [O I], (iii) shock heating produces very high temperatures (of the order of millions of K), which lead to collisional ionization, (iv) this high temperature gas emits X-rays, which contribute to the ionization in front of the shock and behind it, and (v) in the region between the X-ray emitting zone and the low excitation zone, the temperature of the gas passes through intermediate temperatures of several tens of thousands of K, boosting the emission of ultraviolet collisional excited lines. 
The [O III] $\lambda4363$/[O III] $\lambda5007$ ratio can be significantly enhanced. While the bulk of the [O III] $\lambda4363$ emission originates in shocks, the [O III] $\lambda5007$ line is emitted mainly by the photoionized component. Therefore, a large [O III] $\lambda4363$/[O III] $\lambda5007$ ratio ($>0.01$) indicates a significant contribution from shocks.

We have analyzed the [O III] $\lambda5007$/H$\beta$ vs. [O III] $\lambda4363$/[O III] $\lambda5007$ emission-line ratios of our sample galaxies with the help of models combining AGN photoionization and shocks (Moy \& Rocca-Volmerange 2002). These models were computed using CLOUDY (Ferland 1996) and MAPPINGSIII (Dopita \& Sutherland 1996). The emission-line ratios were obtained directly from the SDSS spectra, and therefore correspond to integrated fluxes within a $3\arcsec$ circular aperture (the diameter of the SDSS fiber). 
The results can be seen in Figure~\ref{fig11}. Interestingly, the data can be reproduced only with a mixture of shocks and AGN photoionization.  
All data points lie above log([O III] $\lambda5007$/H$\beta$) = 0.5, and have log([O III] $\lambda4363$/[O III] $\lambda5007$)$>-2.05$. These high ratios can only be accounted for if both photoionization and shocks contribute to the line fluxes (Moy \& Rocca-Volmerange 2002). 
%except for J$1354+1328$ (our control galaxy), whose emission line ratios seem to be well reproduced with only AGN photoionization. 
The two galaxies with the highest kinetic power (J$1055+1520$ and J$1346+5228$) exhibit a larger contribution from shocks to the production of the emission lines. In J$1346+5228$ this is consistent with the fact that [Fe II] emission is often interpreted as indicative of shock excitation (e.g., Moorwood \& Oliva 1988). These results are in good agreement with the hypothesis that the spatially offset features in velocity-offset AGNs are shock-excited by outflows/inflows. 

%Photoionization is not the only process leading to the formation of emission lines. Cooling flows (Cox & Smith 1976, Fabian & Nulsen 1977) and interstellar shocks can also produce them. Shocks are ubiquitous in galaxies, caused by jets (jets associated to brown dwarfs, protostars or massive black holes), winds (stellar winds, protostars, active galactic nuclei) or supersonic turbulence. Dopita & Sutherland (1995, 1996) provide a grid of models of line emission produced by pure shocks. However, the efficiency of ionization by stellar photons is such that, whenever hot stars are present, they are likely to dominate the ionization budget of the surrounding gas. The same can probably be said for radiation arising from accretion onto a massive black hole.

\begin{figure*}
\epsscale{.99}
\plotone{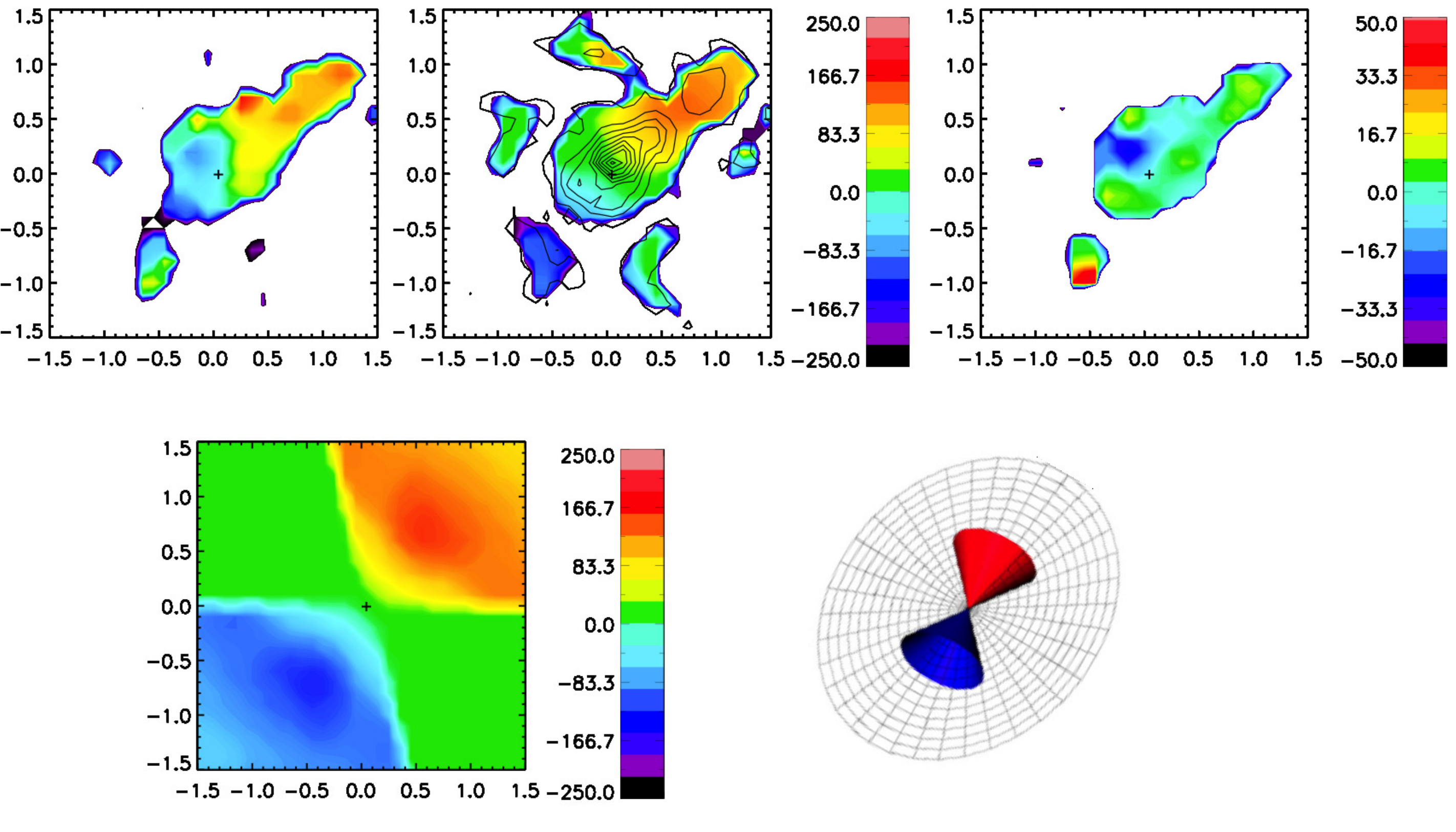}
\caption{Same as Figure 8 but for J$1055+1520$.}
\label{fig9}
\end{figure*}

\begin{figure*}
\epsscale{.99}
\plotone{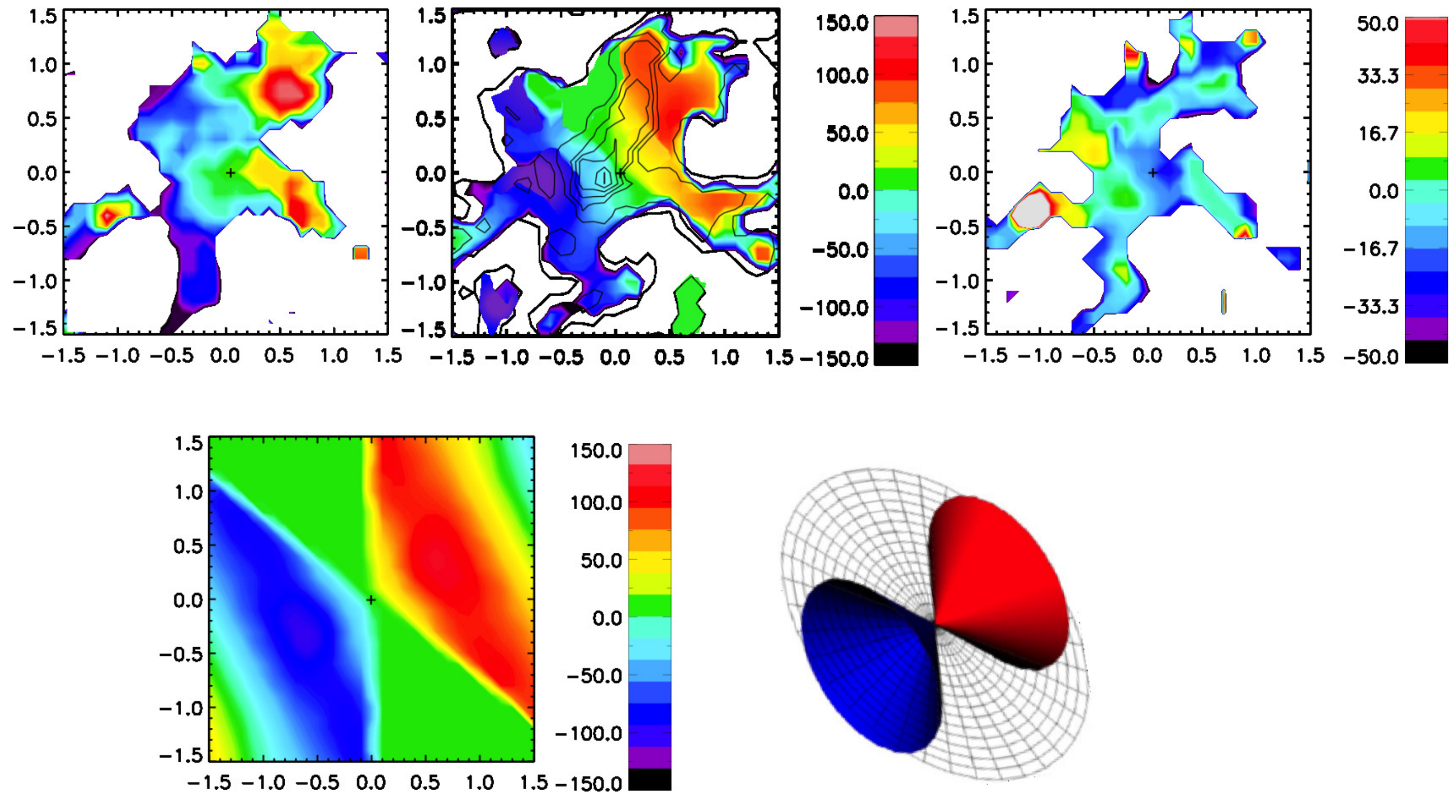}
\caption{Maps showing the data and the best-fit kinematic model used to reproduce the [Fe II] kinematics in J$1346+5228$. The maps are centered at the position of the peak of $J-$band continuum, and the scales are in km s$^{-1}$. \textit{Top Left:} [Fe II] velocity field used for the modeling (same as the one shown in Figure 5). \textit{Top Middle:} Best-fit kinematic model of a biconical outflow with flux contours. \textit{Top Right:} Residuals (data-model) to the fit. \textit{Bottom Left:} Bicone projection in the plane of the sky. \textit{Bottom Right:} Geometric model of the NLR and the inner galactic disk.}
\label{fig10}
\end{figure*}

\subsection{The Integrated Spectral Profiles of the Emission Lines}\label{profiles}

As can be seen in Table 2, the near-IR emission lines in our OSIRIS integrated spectra exhibit the same velocity offset (within the errors) as the optical emission lines in the SDSS spectra. 
%velocity offsets of the emission lines in our OSIRIS spectra are consistent with the offsets measured in the SDSS spectra of the galaxies in our sample. 
However, in all the galaxies that host outflows, the integrated spectral profiles of the emission lines are different. 
%(except in J$1117+6140$, the only galaxy without outflows). 
In these three galaxies, the Pa$\alpha$ emission line ([Fe II] in J$1346+5228$) is asymmetric with blue wings, indicative of disturbed kinematics (Section~\ref{outflows}). In contrast, the optical Balmer lines and [O III] $\lambda5007$ exhibit symmetric profiles (Comerford \& Greene 2014). As an example, we show in Figure 12 the integrated OSIRIS spectrum of J$1055+1520$ in a $3\arcsec$ circular aperture and its optical SDSS spectrum. 

To quantify these asymmetries, we used the weighted skewness ($S$) as described in Comerford \& Greene (2014). $S$ is a popular statistic, defined as the third moment of a distribution function, which describes its asymmetry (see
also Kurk et al. 2004 and Kashikawa et al. 2006). We used the statistical skewness of a line's continuum-subtracted flux,
over the wavelength range where the continuum-subtracted flux values are greater than $10\%$ of the lineÕs peak continuum-subtracted
flux value. Following Comerford \& Greene (2014), nearly symmetric emission lines have $-0.5<S<0.5$. We find that J$1018+2941$, J$1055+1520$ and J$1346+5228$ have $S$ values of $-0.52$, $-0.55$ and $-0.61$, respectively. The minus sign indicates a blueshifted asymmetry, and the typical uncertainty in $S$ is $\pm0.06$. The optical emission lines in the SDSS spectra of these three galaxies have $S$ values between 0 and $-0.5$ (Comerford \& Greene 2014). J$1117+6140$, the only galaxy without outflows, has $S=0.37$, and therefore the redshifted asymmetry is not significant, consistent with its SDSS spectrum. 

There are two possible explanations for this phenomenon. It might be that the near-IR recombination lines are more prone to be formed in regions where shocks dominate (inflows/outflows) due to the low energies involved in moving electrons from lower to higher energy states in the Paschen and Bracket series, while the optical recombination lines require a large contribution from both AGN and stellar photoionization. The second explanation involves obscuration by dust and molecular gas. If part of the outflowing gas is obscured by dust lanes, or if the outflow is optically thick (e.g., Pounds et al. 2003), then it would be difficult to detect the faint blue/red wings in the optical emission lines. The effects of obscuration are much reduced in the near-IR, and therefore we are able to see the true shapes of the spectral profiles.  

%In any case, the asymmetric near-IR profiles of the four velocity-offset AGNs studied in this work do not fulfill the criteria established by Comerford \& Greene (2014) for being offset AGN candidates, but since their selection was done in the optical (where the emission lines are symmetric), they appeared as such. Our observations demonstrate the importance of near-IR integral-field spectroscopy for the study of the gas kinematics in the central regions of galaxies, and indicate that one should be careful when selecting offset AGN candidates from SDSS spectra alone. A near-IR spectroscopic survey similar to SDSS would be a great asset for research programs like this.
%Looking ahead, however, the study of velocity-offset AGNs  is ideally suited for JWST and TMT, because its improved sensitivity and spatial resolution promises to reveal in detail the near-IR properties of the central 50 pc of nearby AGN

%Particularly, in the context of galaxy mergers and offset AGN, the 

%the measured velocity offset of the Pa$\alpha$ emission line ([Fe II] in J) is consistent with the offset of the optical emission lines measured in the SDSS spectrum  

\begin{figure}
\epsscale{.99}
\plotone{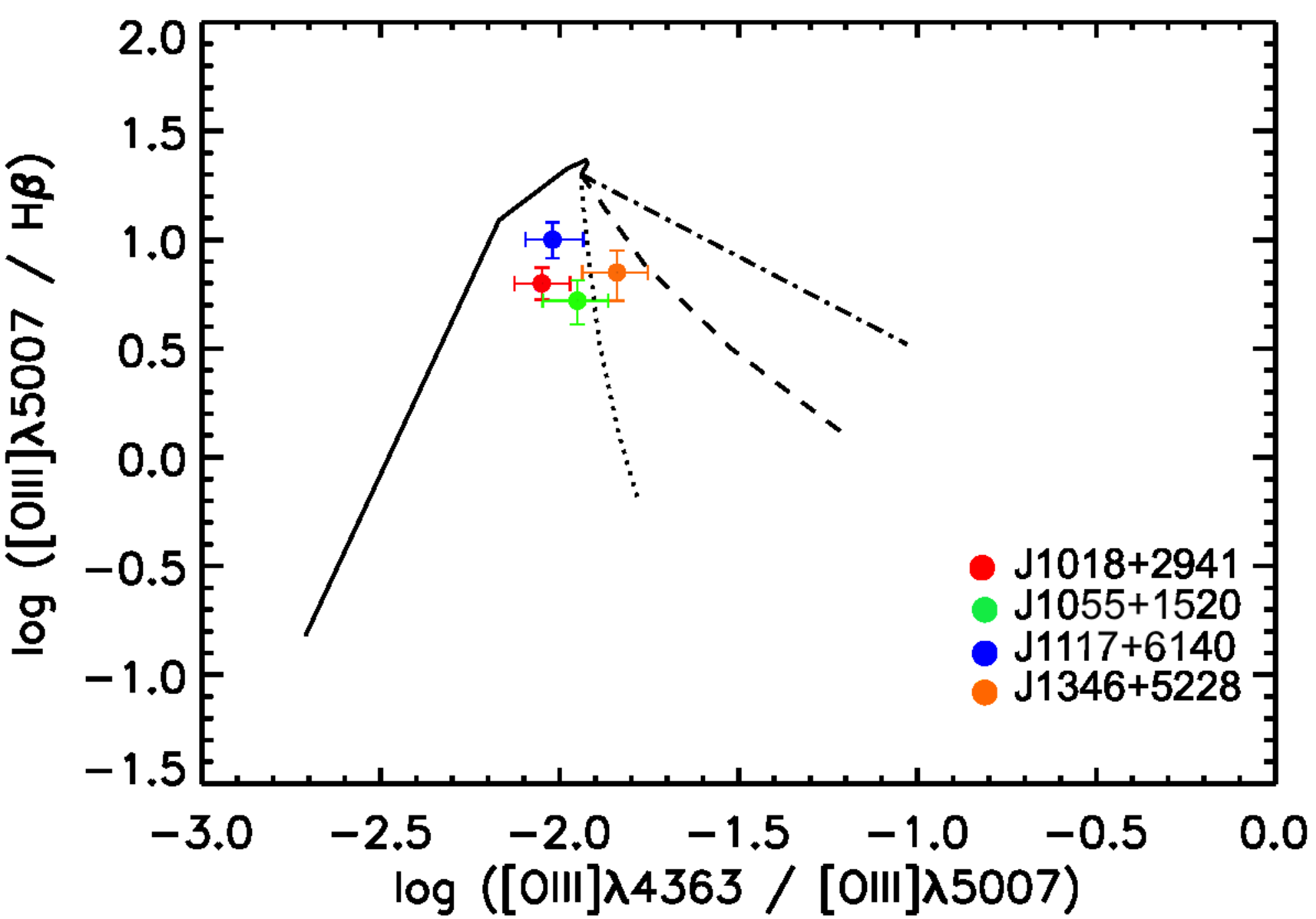}
\caption{[O III] $\lambda5007$/H$\beta$ vs. [O III] $\lambda4363$/[O III] $\lambda5007$. The thick solid line represents pure photoionization models with ionization parameter $U$ varying between log$U=-4$ (bottom) and log$U=-1$ (top), and a spectral index of the ionizing continuum $\alpha=-1$. The dotted, dashed and dotted-dashed lines represent photoionzation $+$ shock models with shock velocities $v=1000$, 300 and 100 km s$^{-1}$, respectively. The metallicity is solar, and the density $N_e$ is equal to 100 cm$^{-3}$. 
The data trends and dispersions are compatible with photoionization plus shock models. Pure photoionization models are clearly excluded %, except for J$1354+1328$ 
(models adapted from Moy \& Rocca-Volmerange 2002).
\label{fig11}}
\end{figure}

\begin{figure*}
\epsscale{.99}
\plotone{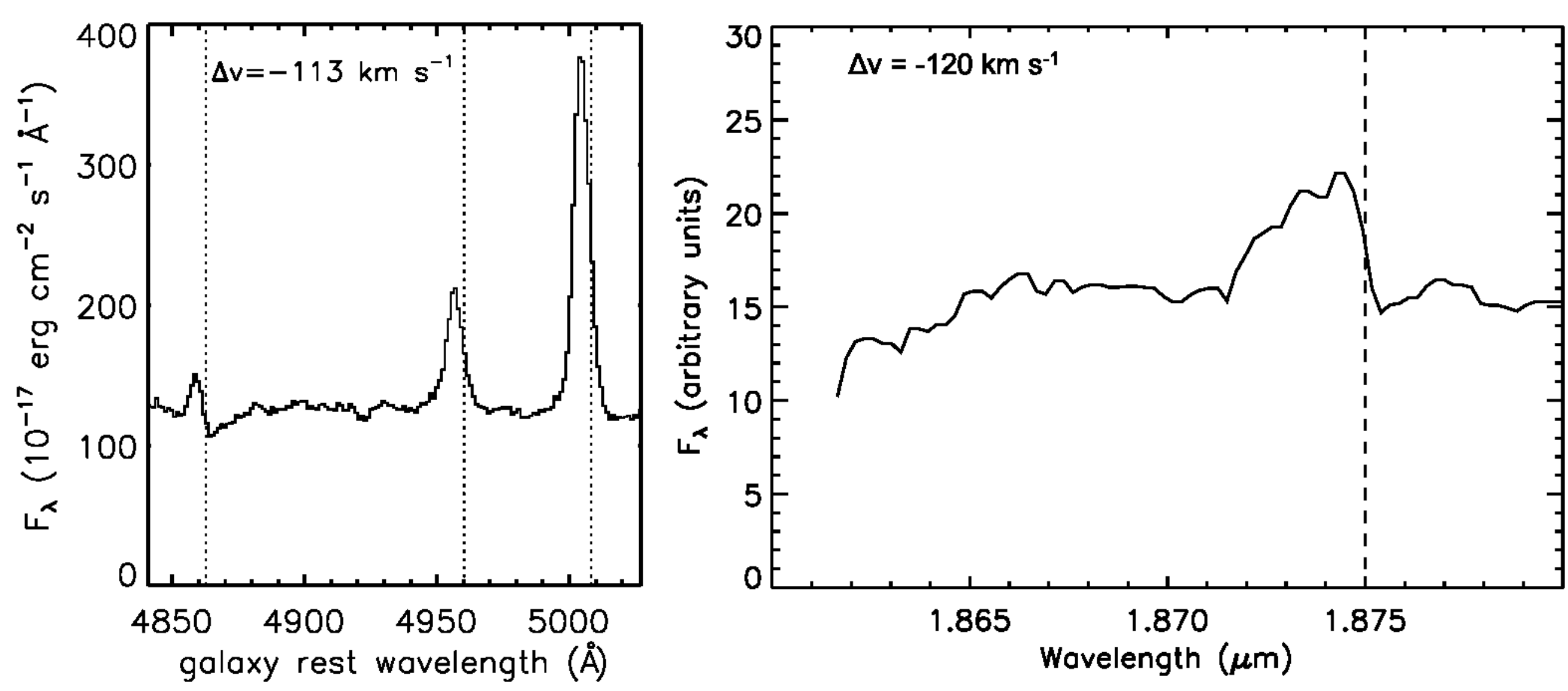}
\caption{Example SDSS (left) and OSIRIS (right) spectra of J$1055+1520$, where each spectrum is shifted to the rest-frame of
the host galaxy's stars. The dotted lines show the rest wavelengths of the stellar absorption features Ca H \& K. In the left panel the spectral profiles of H$\beta$, [O III] $\lambda4959$, and [O III] $\lambda5007$ can be seen with their corresponding velocity offsets (reproduced from Comerford \& Greene 2014). In the right panel the asymmetric profile of Pa$\alpha$ can be seen with its corresponding velocity offset. 
\label{fig12}}
\end{figure*}

\section{Conclusions}

OSIRIS near-IR AO-assisted integral-field spectroscopy has been used in this work to probe the morphology and kinematics
of the ionized gas in four velocity-offset AGNs. %and one control AGN with no velocity offset emission lines. 
Thanks to our high S/N integral-field data, we were able to extract the 2D distribution and kinematics of Pa$\alpha$ (in three objects) and [Fe II] (in one object) in the central kpc of the sample galaxies. Our main results and conclusions can be summarized as follows:

\begin{itemize}
	\item In all cases, the integrated OSIRIS spectra in $3\arcsec$ apertures reproduce the velocity offsets observed in the optical SDSS spectra. %(also in the case when this offset is approximately 0 km s$^{-1}$).
	\item In all of the velocity-offset AGNs, the peak of ionized gas emission is not spatially coincident with the center of the galaxy as traced by the peak of near-IR continuum emission (offset by $0.1-0.7$ kpc). We find that the SDSS velocity offset originates at the location of this peak of emission, and the value of the offset can be directly measured in the velocity maps. 
	%\item In the control AGN J$1354+1328$ (which does not exhibit velocity offsets of the emission lines in its integrated spectrum), the peak of Pa$\alpha$ emission is located at the center of the galaxy. This is the only galaxy that shows ordered rotation in the disk. An extended linear structure of ionized gas which is blueshifted by $\sim70$ km s$^{-1}$ is also observed. The morphology and kinematics of this structure suggest that it is a gas streamer that is emanating from J$1354+1328$ and is traveling towards the companion galaxy J$1354+1327$. 
	\item The spectral offset of the emission lines in the SDSS spectra of J$1018+2941$, J$1055+1520$ and J$1346+5228$ is caused by AGN-driven outflows.
	\item In J$1117+6140$, a counterrotating nuclear disk is observed that contains the peak of Pa$\alpha$ emission $0.2\arcsec$ from the center of the galaxy.  The most plausible explanation for the origin of this spatially and kinematically offset peak is that it is a region of enhanced Pa$\alpha$ emission located at the intersection zone between the nuclear disk and the bar of the galaxy. 
	\item In the galaxies with outflows the mass outflow rates are in the range $70-630$ M$_\odot$ yr$^{-1}$ and the estimated kinetic powers between $2-19\times10^{42}$ erg s$^{-1}$. These are sufficient to create a powerful feedback effect as predicted by theoretical models ($>0.005\, L_{\mathrm{bol}}$, Hopkins \& Elvis 2010). However, no significant differences were found between the outflows produced in velocity-offset AGNs and those present in active galaxies having a near-to-zero-velocity offset in SDSS. 	
%	The nature of this peak is uncertain, with the two possibilities being a region of enhanced Pa$\alpha$ emission located at the intersection zone between the nuclear disk and the bar of the galaxy, or an offset AGN. Follow-up observations with the VLBI would be useful to reveal the nature of this source.
%	\item J$1354+1328$ is the only galaxy that shows clear rotation and an extended linear structure of ionized gas which is blueshifted with velocities of $\sim100$ km/s. The morphology and kinematics of this structure suggest that it is a gas streamer that is emanating from J$1354+1328$ and is traveling towards J$1354+1327$. 
	\item %While the emission line ratios of J$1354+1328$ can be well reproduced by AGN photoionization, 
	The emission-line ratios of the four velocity-offset AGNs in our sample can be reproduced only with a mixture of shocks and AGN photoionization. Shocks provide a natural explanation for the origin of the spatially offset peaks of ionized gas emission observed in these galaxies. 
	\end{itemize}

Our observations demonstrate the importance of near-IR integral-field spectroscopy for the study of the gas kinematics in the central regions of galaxies, and indicate that one should be careful when selecting offset AGN candidates from SDSS spectra alone. The asymmetric near-IR profiles of the four velocity-offset AGNs studied in this work do not fulfill the criteria established by Comerford \& Greene (2014) for being offset AGN candidates, but since their selection was done in the optical (where the emission lines are symmetric), they appeared as such. A near-IR spectroscopic survey similar to SDSS would be a great asset for research programs like this.

\acknowledgments

%We thank the anonymous referee for helpful suggestions that have improved the clarity and strength of this work. We also thank Anil Seth for helpful comments and for providing us with coordinate frame-corrected Hubble Legacy Archive images.

The data presented herein were obtained at the W.M. Keck Observatory, which is operated as a scientific partnership among the California Institute of Technology, the University of California and the National Aeronautics and Space Administration. The Observatory was made possible by the generous financial support of the W.M. Keck Foundation. The authors wish to recognize and acknowledge the very significant cultural role and reverence that the summit of Mauna Kea has always had within the indigenous Hawaiian community. We are most fortunate to have the opportunity to conduct observations from this mountain. The authors thank Hien Tran and Randy Campbell for their support at the W. M. Keck Observatory. We also thank the anonymous referee for helpful suggestions that have improved the clarity and strength of this work. The work of DS was carried out at Jet Propulsion Laboratory, California Institute of Technology, under a contract with NASA. 

%and all those who assisted in the observations at the VLA

Facilities: \facility{Keck:I (OSIRIS), $HST$ (WFC3)}

\end{document}